\documentclass[12pt]{article}
\usepackage{color}
\usepackage{latexsym}
\usepackage{amsmath}
\usepackage{amsfonts}
\usepackage{color}
\usepackage{graphicx}
\usepackage{amssymb}
\usepackage{indentfirst} 
\numberwithin{equation}{section}
\newcommand{\be}{\begin{equation}}
\newcommand{\ee}{\end{equation}}

\newcommand{\mT }{{\mathbb T}}

\newcommand{\mB}{{\mathcal B}}

\newcommand{\tr}{\textrm{tr}}
\newcommand{\mA}{{\mathcal A}}

\newcommand{\bx}{{\bf{x}}}

\title{Evolution of  capacity of entanglement and modular entropy in harmonic chains and  scalar  fields} 
\author{
\\ 
\vspace*{0.3cm}
K. Andrzejewski\footnote{e-mail: krzysztof.andrzejewski@uni.lodz.pl}
 \vspace*{0.3cm}
\\
\small  Faculty of Physics and Applied Informatics  \\ \small
 University of  Lodz\\ \small Pomorska 149/153,
90-236, Lodz, Poland\\
}
\date{}
\begin{document}
\maketitle 
\begin{abstract} 
We examine the  temporal evolution  of the   modular entropy and capacity (in particular, the  fluctuation of the entanglement entropy)  for   systems of  time-dependent    oscillators coupled by  a (time-dependent)  parameter.   Such  models, through the discretization procedure,  fit into   field theory problems  arising from quench phenomena or  non-static spacetimes. 
\par 
First, we compare the  dynamics of   the  modular  and R\'enyi   entropies and derive the form of the modular capacity for the  single time-dependent  oscillator as well as  chains with   bipartite decompositions.   In the latter case  we  analyse  distinguished  periodicities during the evolution and the role  of  various boundary conditions. Next, we focus on the  dynamics  of the  capacity (fluctuation)  of   entanglement. We compare the results obtained  with the predictions of  quasiparticles models;   in particular,  we   obtain  a  theoretical value of the initial slope of the capacity for  abrupt quenches.  We  study also    continuous protocols with  the frequency that vanishes  at  plus (and minus) infinity, including   a model in which the frequency tends to the Dirac delta.      All  the above issues   are discussed   with the emphasis on the analytical methods.
\end{abstract} 
\newpage 
\section{Introduction}
\label{s0}
The notion of the entropy appears 	in many branches of  physics and thus various approaches  and generalizations to it  have been proposed.   The  von Neumann (vN)  entropy  
plays the  prominent role due to its applications    in quantum physics,  especially  in  the  study of the entanglement  which, in turn,  is fundamental  for quantum information processing and technology.  Obviously, the   vN entropy provides only  an overall    characterization  of the state (in particular, the  entanglement of  the state). 
In order to get more information  some other measures, such as the R\'enyi or Tsallis entropies,  were proposed.   The motivation for  various  measures of  the entanglement arises also  from  attempts   to understand  this notion  in  quantum field theory; this  is challenging due to divergences (and thus  the necessity of a regularization procedure, e.g. UV cutoff).   However,  the progress has been made in the recent years, see  e.g. \cite{b1a,b1b,b1c} for review.  One of the most intensively studied  issues  was  the relation  between the entropy and area \cite{b2a,b2}  emerging   from black holes physics (see   \cite{b3a,b3b} and references therein)  and    its   generalization   in the context  gauge/gravity duality (holographic entanglement entropy \cite{b4}).
\par
An interesting  view  on the notion of entropy  can be obtained by considering   a (semipositive) operator  defined as  the minus logarithm of the density operator.  Then, the  vN entropy becomes  the expectation value  of such an operator, which  is  therefore called  the entanglement  (or modular)  Hamiltonian, see,  among others, \cite{b5a}-\cite{b5d}. By having  the  modular Hamiltonian we can apply   thermodynamical  approach and construct   thermodynamical quantities, in particular  the entropy and capacity.   Such an approach leads to new notions, the so-called modular entropy and  modular capacity \cite{b5c,b6a}.  Obviously, these notions  will be useful if they allow us to understand better the  physical phenomena.  For the modular entropy such a situation appears in  the study of  the relation  between the geometry of spacetime and strongly coupled field theories via the AdS/CFT correspondence (in  the Ryu-Takayanagi approach \cite{b4}).   It turns out that    in some cases  the modular entropy  seems  a   natural generalization  of the vN entropy  \cite{b6a,b7a,b7b}.
\par
At the same time, the notion of the modular capacity gained also some attention. Originally it   appears in the  condensed mater physics \cite{b5c}    and   recently in other contexts   (in particular, quantum gravitational effects) \cite{b9a}-\cite{b9h} (for more references see \cite{b9h}). It turns out that,   for  some specified  value of parameter  the modular capacity reduces to the variance of entropy, thereby  it measures the fluctuation  of the entanglement entropy \cite{b10a,b10b}. In this way the capacity of the entanglement  gives another   qualitative measure  of the entanglement spectrum \cite{b11a}-\cite{b11d}. 
 \par
 In this work we analyse   and compare the  modular entropy and capacity   with  the special emphasis put on   their dynamics.   We will focus on the   harmonic chains  with time-dependent parameters as examples of  continuous-variable quantum systems. Such systems are  interesting  due to discretization procedure of  various time-dependent  field models  resulting   from quench phenomena and  non-static  spacetimes.  We start by reviewing  briefly the recent results on this topic and outlining the plan of the paper. Namely, the dynamics of the  vN entanglement entropy of time-dependent  harmonic chains was analysed in various approaches, see \cite{b12a}-\cite{b12h}; despite this effort  some misunderstanding appeared.  We clarify these issues and  then   extend them  to the  modular entropy  and/or  continuous  quenches, see Secs. \ref{s2} and \ref{s3}.   
\par 
On the other hand,  the evolution of the capacity (fluctuation)  of the entanglement  has been recently  analysed for infinite harmonic chains (and the abrupt protocol) in Ref.  \cite{b9h}. Such a model corresponds to  1+1 dimensional  (conformal) field theory on the infinite  line. In particular,  the problem of different slopes  for the linear growth of  capacity and entropy has been indicated.   In Sec. \ref{s4}  we consider  these issues  for the finite size  systems by considering various boundary conditions as well as  the quasiparticles approach.  In Sec. \ref{s5} we study the above points for  continuous  quenches   by means of some analytical examples. 
\section{Preliminaries}
\label{s1}
The  von Neumann (vN)  entropy  of the  density matrix $\rho$ 
\be
S_{vN}= S= -\tr(\rho\ln(\rho)),
\ee
is one of the basic information-theoretic measures.  However, the   vN entropy provides only a  first  characterization of   $\rho$ (e.g. entanglement),  in order to get more information  some other measures were proposed.  One of them is the R\'enyi  entropy defined  as
\be 
R_\alpha=\frac{1}{(1-\alpha)}\ln(\tr(\rho^\alpha)),
\ee
  for which   $\lim_{\alpha \rightarrow  1} R_\alpha=S$ holds.  Another  generalization,  motivated by the fact that  trace of the power of density matrix can be used to characterize its  eigenvalues, is given by the Tsallis entropy  $T_\alpha=(\tr(\rho^\alpha)-1)/(1-\alpha)$.
 \par 
In order to get more insight into  the notion of the  entropy let us consider   the logarithm of the density matrix
\be
K= -\ln(\rho).
\ee
Then, the  vN entropy is the expectation value of $K$,   $S=\tr(\rho K)=\langle K\rangle $.  Since    $K$ is semipositively  defined,  $K\geq 0$,  it  is called the modular Hamiltonian (due to the relation $\rho=\exp(-K)$). For  the density matrix  describing  a part of  a bipartite system (i.e. the reduced density)  the vN entropy  becomes the  basic measure of the entanglement;  thus sometimes  the modular Hamiltonian is called the entanglement  Hamiltonian  or entropy operator (such a nomenclature appears especially  in the context of the condensed matter physics). 
\par 
 Following the above  observation we can consider analogues of thermodynamical  quantities with respect to $K$; in particular,   the notion of entropy for   $K$. 
 To this end  we denote  the inverse   $\alpha=1/T$  of  ``temperature" $T$  (we use $\alpha$ instead of thermodynamical  $\beta$ index and the Boltzmann  constant  is put   one)    and  we define  the partition function
 \be
 Z_\alpha=\tr(e^{-\alpha K})=\tr(\rho^\alpha) ,
 \ee
 as well as the corresponding  free energy 
 \be
 F_{\frac{1}{T}}=F_\alpha= -\frac{1}{\alpha} \ln(Z_\alpha).
 \ee
 Then,   following  the standard thermodynamical  approach we construct the corresponding entropy  
 \be
S_\alpha=-\partial _T F_\frac 1 T=\alpha^2\partial_\alpha F_\alpha.
 \ee
The new quantity $S_\alpha$  is called the modular entropy (due to its  relation to the modular Hamiltonian $K$)  and      can be easily compared  with  the  R\'enyi entropy 
 \be
 \label{ee1}
 S_\alpha=\alpha^2\partial_\alpha(\frac{\alpha-1}{\alpha}R_\alpha)=R_\alpha+\alpha(\alpha-1)\partial_\alpha R_{\alpha}.
 \ee
Equivalently,  it is  the vN entropy of the modified density operator $\rho_\alpha=\rho^\alpha/Z_\alpha$, i.e. $S_\alpha=\langle K\rangle_\alpha$  where $\langle \cdot \rangle _\alpha $  stands for the expectation value at $\rho_\alpha$. The last observation gives immediately $S_\alpha\geq 0$ and      $\lim_{\alpha \rightarrow  1} S_\alpha=S_{vN}$.    Some more properties of the  modular  entropy   as well  as its  meaning   in the  context of  gravitational     holography can be found in  Refs.  \cite{b6a,b7a,b7b}. 
 \par The above thermodynamical analogy can be continued  by introducing the   (modular) capacity
$ C_\alpha=-T\partial^2_TF_{\frac 1 T}=-\alpha\partial_\alpha  S_\alpha  $.   Then one gets 
\be
\label{ee3}
C_\alpha=\alpha^2\partial_\alpha^2  ((1-\alpha) R_\alpha)=\alpha^2(\langle K^2\rangle _\alpha-\langle K\rangle _\alpha^2)\geq 0,
\ee
i.e. $C_\alpha$ measures the variance of the modular Hamiltonian at $\rho_\alpha$. Since  for  $\alpha=1$ the  modular entropy coincides with the vN one ($S_1=S_{vN}$),  the  capacity reduces to the  variance of  the modular Hamiltonian  $C\equiv C_1=\langle K^2\rangle -\langle K\rangle ^2$, see \cite{b10a}-\cite{b11d}. In consequence, it gives the quantum fluctuation with  respect to the original state.  For the reduced density  matrix  $C$   measures  the  entanglement entropy fluctuations.    The magnitude of the relative entanglement fluctuations is   defined as $\delta K=\sqrt C/S$ \cite{b10a,b9a}.
As a  border between strong and weak fluctuations  the condition $\delta K=1$ can be used,   i.e. $C=S^2$. Similarly, for an arbitrary  $\alpha$   we have 
\be
\langle K \rangle _\alpha=\partial_\alpha ((\alpha -1)R_\alpha ) ,
\ee
and  we can  also  easily express    $\delta_\alpha K=\sqrt{C_\alpha}/(\alpha \langle K\rangle _{\alpha})$   by means of the R\'enyi entropy.
\par 
Concluding, let us  note that  for a finite-dimensional  system   and  the  density matrix  proportional to the  identity  (e.g. for the  maximally entanglement states)  all the  entropies  $S,R_\alpha,S_\alpha $ (for any $\alpha$)  coincide and $C_\alpha=\delta_\alpha K=0 $.   On the other hand, the maximal  value  of the capacity appears for partially mixed states \cite{b10a,b9a}.   For the continuous variables systems the situation is  more complicated and will  be discussed below.

\section{Warm-up: time-dependent oscillator}
\label{s2}
In this section we examine the temporal  evolution of the above quantities in a very simple case. Namely,
let us take the harmonic oscillator with the time-dependent frequency 	$\omega(t)$. It is described  by the  Hamiltonian
\be
\label{e1}
H(t)=\frac{p^2}{2}+\frac{1}{2}\omega^2(t)x^2,
\ee
for which the  equation of motion reads
\be
\label{e2}
\overset{..}{x}(t)=-\omega^2(t)x(t).
\ee
 The time-dependent harmonic oscillator (TDHO)  appears in many physical models and has been studied in various contexts.  It turns out  that  eq. \eqref{e2}   is equivalent  to   the so-called  Ermakov-Milne-Pinney (EMP) equation \cite{b14a,b14b,b14c} 
 \be
 \label{e3}
\overset{..}{b}(t)+\omega^2(t)b(t)=\frac{c^2}{b^3(t)},
\ee 
where $c$ is a constant (we assume  $c\neq 0 $ to ensure the non-vanishing of the function $b(t)$). 
At the quantum level,   the general solution to  the Schr\"odinger equation of the TDHO is  a superposition of the following wave functions 
\be
\label{e11}
\psi_n(x,t)=\frac{1}{\sqrt{2^nn! b(t)}}\sqrt[4]{\frac{c}{\pi}}e^{-ic(n+1/2)\tau(t)}H_n\left(\frac{\sqrt{c} x}{ b(t)}\right)e^{-\frac{cx^2}{2b^2(t)}+\frac{i\dot b(t)x^2}{2b(t)}},
\ee
where $H_n$ for $n=0,1\ldots$ denote the Hermite polynomials and $\tau(t)=\int b^{-2}(t)$.
 \par
 In  physical considerations  we usually assume that  the states \eqref{e11} at   initial time $t=t_0$ are   eigenstates  of the instantaneous Hamiltonian $ H(t_0)$.  This holds for the following  initial conditions 
\be
\label{e15}
 c=\omega(t_0)\equiv\omega_0,\quad  b(t_0)=1, \quad \dot b(t_0)=0,
\ee
and  will be assumed in what follows. 
 \par 
   Then  the density function of the state $\psi_n(x,t)$  reads 
\be
\label{e37}
\rho_n(x,t)=\frac{1}{2^n n! b(t)}\sqrt{\frac{\omega_0}{\pi}}H_n^2\left(\frac{x\sqrt{\omega_0}}{b(t)}\right)e^{-\frac{\omega_0 x^2}{b^2(t)}},
\ee
while the   density of the  Fourier transform of $\psi_n(x,t)$ is given by the formula 
\be
\label{e38}
 \rho_n(p,t)= \frac{b(t)}{ 2^n n!\sqrt{\omega_0^2+b^2(t)\dot b^2(t)}}\sqrt{\frac{\omega_0}{\pi}}e^{-\frac{\omega_0b^2(t)p^2}{\omega_0^2+b^2(t)\dot b^2(t)}}H_n^2\left(\frac{\sqrt{\omega_0}b(t)p}{\sqrt{\omega_0^2+b^2(t)\dot b^2(t)}}\right).
\ee
 We start with  the  coordinate     R\'enyi entropy
\be
\label{e42}
R^{\alpha,x}_n(t)=\frac{1}{1-\alpha}\ln\left (\int \rho^\alpha_n(x,t) dx\right),
 \ee 
 as well as  with  their  momentum   counterpart $R^{\alpha,p}_n(t)$. 
 For the  TDHO they read respectively  
\be
\label{e42b}
\begin{split}
 R_{\alpha,n}^{x} (t)&=\ln\left(\frac{b(t)}{ \sqrt{\omega_0}}\right)+\frac{1}{1-\alpha}\ln(W_{\alpha,n}),\\
R_{\alpha,n}^{p}(t)&= \frac 1 2 \ln\left(\frac{\omega_0^2+b^2(t)\dot b^2(t)}{\omega_0^2b^2(t)}\right) +\frac{1}{1-\alpha}\ln(W_{\alpha,n}),
\end{split}
\ee 
where  $W_{\alpha,n}= \int_{-\infty}^{\infty}  \left(\frac{H_n^2(y) }{\sqrt \pi 2^n n!}\right)^\alpha  e^{-\alpha y^2}dy$ is the so-called entropic moment of the Hermite polynomials  \cite{b13x}.
  For  lower states the latter quantities  can be given  explicitly 
\be
W_{\alpha,0}=\sqrt{\frac {\pi^{1-\alpha}}{\alpha}},\quad  W_{\alpha,1}=\frac{2^\alpha}{\pi^{\alpha/2}\alpha^{\alpha+1/2}}\Gamma(\alpha+1/2);
\ee
for  higher $n$ they are a  more  complicated  combination of the Bell polynomials \cite{b13x}.
By virtue of  eq. \eqref{ee1}  we conclude  that the  coordinate  modular entropy is given by  
\be
\label{ee2}
S_{\alpha,n}^x(t)= R_{\alpha,n}^{x} (t)+\frac{\alpha}{\alpha-1}\ln(W_{\alpha,n})-\alpha\partial_\alpha\ln(W_{\alpha,n}),
\ee
and analogously for the momentum case. From eq. \eqref{ee2}   we   see that the time dependence of the    modular entropies $S_{\alpha,n}$ is the same as for the   R\'enyi ones; in particular,  the increase  of  the modular  entropy for an arbitrary state $\psi_n$  does not depend on $n$, e.g.  $S_{\alpha,n}^{x} (t)-S_{\alpha,n}^{x} (t_0)=\ln(b(t))=R_{\alpha,n}^{x} (t)-R_{\alpha,n}^{x} (t_0)$ (obviously it vanishes for the ordinary harmonic oscillator, $b(t)=1$). In summary,  for the  states  $\psi_n(t)$   both kinds of  entropies differ by a  constant (depending on $\alpha$ and $n$) only. 
\par 
Now, let us  pass  to the notion of the  modular capacity. By virtue of eq. \eqref{ee3}  we have
\be
\label{ee11}
C_{\alpha,n}^x(t)=C_{\alpha,n}^p(t)=\alpha^2\partial^2_\alpha \ln(W_{\alpha,n})\equiv C_{\alpha,n}(t);
\ee
thus the capacity   is time-independent (and bounded  for fixed $\alpha$ and $n$). For the lowest (initial)   state we have  for any $\alpha$
\be
C_{\alpha,0}(t)=\frac 1 2 ,
\ee
while for the excited state the $\alpha$-dependence emerges
\be
C_{\alpha,1}(t)=\frac  1 2-\alpha +\alpha^2 \Upsilon _1(\alpha+\frac 1 2),
\ee
where  $\Upsilon _1$ is trigamma function (the first derivative of the digamma function).
For the most interesting case $\alpha=1$   (corresponding to the fluctuation of  the vN entropy)   we have
\be
\label{ee15}
C_{1,1}(t)=\frac 1 2 (\pi^2-9).
\ee
For   $n\geq 2$    only numerical results  can be obtained.  Finally,   let us note that the above considerations   can be extended to  the  TDHO  driven by an external time-dependent force, i.e. 
\be
\label{er4}
\overset{..}{x}(t)=-\omega^2(t)x(t)+f(t).
\ee
 Namely, after direct   calculations,  we find that   all the  entropies   reduce to the ones for the force-free case.
 \par
 In summary,   in this section  we  showed  that  the dynamics of  the modular entropy  of  the  basic solutions of the  TDHO is the same  as for  the R\'enyi entropy.   This supports the point of view   that the modular entropy  can be considered  as another candidate for  a  generalization of the   vN entropy.   Moreover, we showed   that  the capacity (fluctuation) of the vN entropy for such    solutions of the  TDHO is time-independent and we found  its   form.    In  the next section we will see that the  situation changes when we consider more  complicated  systems   consisting  of   more  oscillators and reduced densities.
\section{Time-dependent coupled oscillators}
\label{s3}
In this section we  investigate the   notion of the modular entropy and  the capacity in the context of entanglement  of  the systems. To this end  we  consider  the system of  TDHO coupled by  a time-dependent  parameter and next we  analyse various bipartite  decompositions and corresponding reduced density operators. 
\subsection{Two coupled oscillators}
\label{s3a}
Let us consider the Hamiltonian of  the form  
\be
\label{e81}
H(t)=\frac 12(p_1^2+p_2^2)+\frac 12\omega^2(t)\left((x^1)^2+(x^2)^2\right)+\frac 12k(t)\left(x^1-x^2\right)^2.
\ee
By means of    the transformation ${x}=R{ y}$ where $R$  is the 2-dimensional  rotation matrix  with the rotation  angle $ \pi/4$, we transform  the Hamiltonian \eqref{e81} into the following one
\be
\label{e82}
H_{y}(t)=\frac 12(p_1^2+p_2^2)+\frac 12\left(\omega^2_1(t)(y^1)^2+\omega^2_2(t)(y^2)^2\right),
\ee
where  now  $p$'s denote  the canonical momenta associated  with $y$'s and $\omega_1^2(t)=\omega^2(t)+2k(t),\quad \omega_2^2(t)=\omega^2(t)$.    The frequencies $\omega_{1,2}(t)$ determine the  parameters of the initial Hamiltonian  \eqref{e81} as follows 
\be
\label{e83}
\omega(t)=\omega_2(t),\quad k(t)=\frac 12 (\omega_1^2(t)-\omega_2^2(t)).
\ee
\par 
The evolution  $\psi_0(x,t)$  of the ground state  $\psi_0(x,t_0)$ of the Hamiltonian  operator  $ H(t_0)$ as well as the  reduced density matrix $\rho_0^{red}(x^1,\tilde x^1,t)=\int \psi_0(x^1,x^2,t)\psi^*_0(\tilde x^1,x^2,t) dx_2$    can be  easily computed    when we  take into account the form of the  Hamiltonian  \eqref{e82} and then  return to the $x=(x^1,x^2)$ variable. 
The final result reads  \cite{b12b}
\be
\rho_{0}^{red}(x^1,\tilde x^1,t)=\sqrt{\frac{(\zeta(t)-\chi(t))}{\pi}}e^{\chi(t)x^1\tilde x^1+ i((x^1)^2-(\tilde x^1)^2)\varphi(t)-\frac{\zeta(t)}{2}((x^1)^2+(\tilde x^1)^2)},
\ee
where 
\be
\label{e85a}
\varphi(t) =\frac{\dot b_1(t)}{4b_1(t)}+\frac{\dot b_2(t)}{4b_2(t)}-\frac{\frac{c_1}{b_1^2(t)}-\frac{c_2}{b_2^2(t)}}{\frac{c_1}{b_1^2(t)}+\frac{c_2}{b_2^2(t)}}\left(\frac{\dot b_1(t)}{4b_1(t)}-\frac{\dot b_2(t)}{4b_2(t)}\right),
\ee
 and $\zeta(t)>\chi(t)\geq 0$ are given by 
\be
\label{e85}
\zeta(t)=\frac{\left(\frac{c_1}{b_1^2(t)}+\frac{c_2}{b_2^2(t)}\right)^2+4\frac{c_1c_2}{b_1^2(t)b_2^2(t)}+\left(\frac{\dot b_1(t)}{b_1^2(t)}-\frac{\dot b_2(t)}{b_2^2(t)}\right)^2}{4\left(\frac{c_1}{b_1^2(t)}+\frac{c_2}{b_2^2(t)}\right)},
\quad 
\chi(t) =\frac{\left(\frac{c_1}{b_1^2(t)}-\frac{c_2}{b_2^2(t)}\right)^2+\left(\frac{\dot b_1(t)}{b_1^2(t)}-\frac{\dot b_2(t)}{b_2^2(t)}\right)^2}{4\left(\frac{c_1}{b_1^2(t)}+\frac{c_2}{b_2^2(t)}\right)},
\ee
while  the functions $b_{1,2}(t)$ satisfy the EMP equation \eqref{e3} with the frequencies $\omega_{1,2}(t)$ and the constants  $c_1=\sqrt{\omega^2(t_0)+2k(t_0)}$ and $c_2=\omega(t_0)$, respectively.   
In consequence,  the R\'enyi  entropy  and  the  vN   entropy     of the  reduced density  $\rho_{0}^{red}$  can be easily found,   see   Ref. \cite{b12b}
\be
\label{e84}
R_\alpha(t)=\frac{1}{1-\alpha}\ln\frac{(1-\xi (t))^\alpha }{1-\xi^\alpha(t)}, \quad S(t)=-\ln(1-\xi(t))-\frac{\xi(t)}{1-\xi(t)}\ln\xi(t),
\ee
where $\xi(t)=\frac{\chi(t)}{\zeta  (t)+\sqrt{\zeta^2(t)-\chi^2(t)}}$.
\par
Now, we are in the position to find the modular entropy and capacity. First, the  straightforward computations give  
\be
\label{ee5}
S_\alpha(t)=-\ln(1-\xi^\alpha (t))-\frac{\xi^\alpha(t)}{1-\xi^\alpha(t)}\ln(\xi^\alpha(t)),
\ee 
i.e. $S_\alpha$ is obtained by the replacement  $\xi(t)\rightarrow \xi^\alpha(t)$ in the vN entropy, cf. eq. \eqref{e84}  (this can be also seen from the fact   the spectrum of $\rho_{0}^{red}$ consists of the powers of $\xi$). 
Moreover,  we have the following relation  
\be
S_\alpha(t)=R_\alpha(t)+\frac{\alpha}{\alpha-1}\left(\ln\frac{1-\xi(t)}{1-\xi^\alpha(t)}+\frac{(1-\alpha)\xi^\alpha(t)\ln(\xi(t))}{1-\xi^\alpha(t)}\right);
\ee
thus, in contrast to the  single TDHO, the dynamics of the modular entropy for the coupled  oscillators  is different from   the R\'enyi one. 
\par
Similarly,  for the capacity  we get
\be
\label{ee6}
C_\alpha (t)=\frac{\xi^\alpha(t)\ln^2(\xi^\alpha(t))}{(1-\xi^\alpha(t))^2};
\ee
in consequence,  we have  $0\leq C_\alpha(t)\leq 1$. 
In particular, the capacity (fluctuation)  of the entanglement  is given by the formula
\be
C(t)=C_1(t)= \frac{\xi(t)\ln^2(\xi(t))}{(1-\xi(t))^2}.
\ee
From the above we see that the discussed quantities are  determined by the solutions of the   two EMP equations with the  frequencies $\omega_{1,2}(t)$, respectively. 
\par
In order to analyse the above results, let us start with the simplest case,   namely constant  both   the coupling $k(t)=k$  and the frequency $\omega(t)=\omega$ in the Hamiltonian \eqref{e81}.  Then  $b_1(t)=b_2(t)=1$ and   $\xi(t)=\xi$   is a  constant which   can be easily expressed  as a   function of the ratio $k/\omega^2$ (the same concerns  $S_\alpha,C_\alpha$); in particular,   $\xi\rightarrow 1$ for $k\rightarrow \infty$.    
In this case, the reduced  matrix $\rho_{0}^{red}$  is equivalent to  the thermal density  operator  for a 
single harmonic oscillator specified by the  frequency $ \sqrt{\zeta^2-\chi^2}$   and temperature   
$T=-\sqrt{\zeta^2-\chi^2}/\ln(\xi)$.      The modular entropy is also the vN entropy  of the harmonic oscillator  provided we  
define the  temperature $T=- \sqrt{\zeta^2-\chi^2}/(\alpha \ln(\xi))$.   Moreover, the capacity becomes the  ordinary heat capacity of   a quantum harmonic oscillator.  
Thus  when   $\xi$ tends to one ($k$ to infinity) the  entanglement fluctuations are close to one  while   the entanglement entropy tends to infinity; this is in contrast with the   finite-dimensional  case where the capacity vanishes for the maximally entanglement states.  
\par
For more coupled oscillators the  reduced matrix is not (in general)  equivalent to a thermal density matrix for a system of oscillators \cite{b2}. 
In what follows we will consider  such systems in the more general time-dependent  case.  Then all entropies are time-dependent and measure the  evolution of entanglement  (non-local correlations). This is interesting  from the thermodynamical point of view since   such  models can be considered as many-body non-equilibrium systems. Although the   unqualified   definition of the entropy for non-equilibrium systems is still problematic \cite{R6}  such information  can give some insight into how  thermodynamics arises in isolated quantum systems as well as   probe thermalization processes, see e.g.  \cite{R2,R3,R4} and  \cite{R5} for experimental analysis.  Moreover, such considerations are  relevant  for quantum information processing and technology and can lead to some universal scaling properties as for  the correlations functions in the   Kibble-Zurek scenario \cite{R17}.  
\par
 A  typical example of such a situation  are  many-body quantum  quench phenomena characterized by  sudden  (fast)  changes  of   global  parameters of  the Hamiltonian.   Then it turns out that the entanglement entropy starts to grow linearly in time. Next it is  saturated  to a volume-law behavior suggesting in this way the thermalization of the subsystem, see \cite{b5b,R9} and references therein.   A deeper analysis of such a scenario (see e.g.  \cite{R10,R11,R12})  implies  that the resulting reduced density matrix for a  subsystem relaxes to the Gibbs ensemble or in the case of integrable systems, a generalized Gibbs ensemble. Moreover, the   other approaches  to this problem  are also related to the entanglement entropy.  For example, holographic considerations presented in  Ref. \cite{R13} suggest that entanglement entropy  can be treated also as a kind  of coarse-grained entropy for time-dependent system. Similar conclusions, but  in a different approach \cite{R15},   appear for the  cosmological  spacetimes and non-equilibrium processes.  
\subsection{Many coupled oscillators}
\label{s3b}
The system of $N$ coupled time-dependent   oscillators (with the nearest neighbour interaction)  is described by the Hamiltonian 
\be
\label{ee8}
H(t)=\frac 12\sum_{l=1}^{N}(p_l^2+\omega^2(t)(x^l)^2)+ \frac12k(t)\sum_{l=0}^{N}(x^l-x^{l+1})^2.
\ee
Imposing  boundary conditions  the Hamiltonian can be rewritten in the form 
\be
H(t)=\frac 1 2\sum_{j=1}^{N}p_j^2+\frac 1 2x^T \Lambda (t)x,
\ee
where $x=(x^1,\ldots,x^N)$ and $\Lambda(t)$ is a  symmetric $N\times N$  matrix with  the eigenvalues  $\lambda_j(t)$, for  $j=1,\ldots,N$. 
 The most popular boundary conditions  are the following three:
\begin{itemize}
\item  The periodic  boundary conditions (PBC)  defined by  $x^0=x^1$ and   $x^{N+1}=x^0$.   Then 
 the eigenvalues of $\Lambda(t)$  are  of the form   $\lambda_j(t)=\omega^2(t)+4k(t)\sin^2(\frac {j\pi}{N})$, $j=1,\ldots,N$.  
\item The  Neaumann  boundary conditions (NBC)    given by  $x^0=x^1$ and  $x^{N+1}=x^{N}$ (corresponding to the system of  coupled pendulums  in small-angle approximation); then  the eigenvalues read  $\lambda_j(t)=\omega^2(t)+4k(t)\sin^2(\frac {j \pi}{2N})$, $j=0,\ldots,N-1$ (for  $N=2$  we obtain the previous case, see Sec. \ref{s3a}). 
\item   Finally, we can impose   the Dirichlet boundary conditions (DBC),  $x^{0}=x^{N+1}=0$, then $\lambda_j(t)=\omega^2(t)+4k(t)\cos^2(\frac {j \pi}{2(N+1)})$, $j=1,\ldots,N$.
\end{itemize} 
\par
As in  the case of two oscillators,  we can easily  find  the evolution of the  density matrix of the   ground state  of the instantaneous Hamiltonian  $ H(t_0)$. However, in contrast to the case of two oscillators,  the reduced density for more oscillators   is  substantial  more complicated and   some inconsistency appeared in the literature.  In view of  this, first,   we   discuss this  problem. 
\par 
We start with the (time-dependent) density matrix of the whole  system, see e.g.  \cite{b12b,b13}   
\be
\rho(x,x',t)=\sqrt{\det(\Omega(t)/\pi)}\exp( ix^TB(t) x-i{x'}^TB(t)x'- \frac 12 x^T\Omega(t) x- \frac 12 {x'}^T\Omega(t) x'),
\ee
where $x=(x^1,\ldots,x^N)$ and $\Omega(t)=U^T\sqrt{\tilde\Lambda(t)}U$, $B=U^T\tilde B(t)U$  where $\tilde B(t),\tilde \Lambda(t)$  are diagonal matrices  with elements   $(\tilde\Lambda(t))_{ij}=\lambda_i(t_0)/b_i^4(t)\delta_{ij} $ and $(\tilde B(t))_{ij}=\dot b_i(t)/(2b_i(t))\delta_{ij}$, respectively,  while   $b_j(t)$ are the  solutions  of the EMP equations with the frequencies $\lambda_ j(t)$ 
 \be
 \label{ee9}
\overset{..}{b_j}(t)+\lambda_j(t)b_j(t)=\frac{\lambda_j(t_0)}{b^3_j(t)}, \quad j=1,\ldots,N;
\ee
 and, finally,  $U$ is a time-independent matrix   diagonalizing $\Lambda(t)$: 
 \be
 \label{ee19}
 U\Lambda(t) U^T=Diag(\lambda_1(t),\ldots, \lambda_N(t)).
 \ee
Next, we  split  the whole system into two parts: the first one  $\mathcal{A}$  consisting of   $n$ first oscillators  $(x^1,\ldots, x^n)$   and  the second one $\mathcal{B}$  consisting of  the remaining  $N-n$ ones described by  $\bx=(x^{n+1},\ldots,x^{N})$ and rewrite $\Omega$ and $B$ in the form\footnote{ For simplicity of notation  we omit   the  time parameter $t$ in the matrices when it does not cause ambiguity.} 
\be
\Omega=
\left(
\begin{array}{cc}
\Omega_1&\Omega_2\\
\Omega_2^T&\Omega_3
\end{array}
\right),\quad B=
\left(
\begin{array}{cc}
B_1&B_2\\
B_2^T&B_3
\end{array}
\right),
\ee
where $\Omega_1,B_1$ are $n\times n$ matrices. 
\par  Tracing  (integrating) over the subsystem $\mathcal{A}$ we obtain the reduced density of the subsystem $\mB$. Namely, after straightforward  computations we arrive at the   formula
\be
\rho_{\mathcal {B}}(\bx,\bx',t)= A\exp(i\bx^TZ\bx-i{\bx'}^TZ\bx'-\frac 12 \bx^T\Upsilon \bx -\frac 12{\bx'}^T\Upsilon \bx' +\bx^T\Delta\bx'),
\ee
where $Z,\Upsilon ,\Delta$ are $(N-n)\times (N-n)$ matrices  given by 
\begin{align}
Z&=B_3-B_2^T\Omega_1^{-1}\Omega_2,\\
\Upsilon &=\Omega_3-\frac 12 \Omega_2^T\Omega_1^{-1}\Omega_2+2B_2^T\Omega_{1}^{-1}B_2,\\
\Delta&= \frac 1 2 \Omega_2^T\Omega_1^{-1} \Omega_2+2B_2^T\Omega_1^{-1}B_2+i\Theta,
\end{align}
with $\Theta=\Omega_2^T\Omega_1^{-1}B_2-B_2^T\Omega_1^{-1}\Omega_2$.   Such a form  of the reduced density  coincides with the  one obtained in Ref. \cite{b13}. 
The matrices  $\Omega_1,B_1$  are real and symmetric thus  $\Theta$ is skew-symmetric. In consequence,  for $N=2$   (and  $n=1$) $\Theta$  vanishes  and we obtain the results from  previous section for two oscillators. However, for higher $N$   (and $ n$)   $\Delta$ is a complex (but Hermitian)  matrix.  In consequence, we cannot directly apply  methods  from Ref. \cite{b2}  to obtain the spectrum of $\rho_{\mB}$ 
(i.e.  simultaneously diagonalize both $\Upsilon $   and $\Delta$ by means of an orthogonal matrix).  The  term with $\Theta$  has   been missed   in Refs. \cite{b12b,b12g} leading  to some  problems,   e.g., despite the fact the  initial state is pure the corresponding entropies $S_{\mathcal A} $ and $ S_{\mathcal B}$ were not equal  (this is also in contrast to  results of  Ref.   \cite{b12f}).
  For  similar reasons an  approach in Ref. \cite{b12a} seems  also to miss something; namely,  in our notation the  matrix $\mathcal{E}$ therein  takes the form $\Upsilon^{-1}\Delta$  and thus  has a real (not complex as in  \cite{b12a})  spectrum (in fact the same as the  Hermitian matrix $\tilde \Delta $ below, eq. \eqref{ee21}); moreover,  the matrices    $\mathcal{E} $ and its complex conjugate are, in general,  not  simultaneously diagonalizable and thus  an extension of   the reasoning  from  Ref. \cite{b13b} is more involved. The  problem of the  spectrum of the  density  operator in the case of  Hermitian, but not real, $\Delta$   has been recently discussed,  in the context of squeezed state,  in Ref. \cite{b15b}.   Here we adopt   essential  results and     for more details we refer  to \cite{b15b}. 
\par 
Namely, it turns out that  the spectrum of  the reduced density matrix with  Hermitian  $\Delta$    is of the form 
\be
\label{ee20}
(1-\xi_1)(1-\xi_2)\ldots(1-\xi_{N-n})\xi_1^{m_1}\xi_2^{m_2}\ldots \xi_{N-n}^{m_{N-n}},
\ee
where $\xi$'s are the inverse of the  eigenvalues  (larger than one)  of the following matrix
\be
\left(
\begin{array}{cc}
2\tilde \Delta^{-1}&-\tilde \Delta^{-1}\tilde \Delta^T\\
I & 0
\end{array}
\right),
\ee
where 
\be
\label{ee21}
\tilde \Delta=(\tilde\Upsilon)^{-1/2}\Pi \Delta \Pi^T  (\tilde\Upsilon)^{-1/2},
\ee
 while $\Pi$ is  an orthogonal matrix   diagonalizing  $\Upsilon$, i.e.  $\Pi\Upsilon\Pi^T=\tilde \Upsilon$; let us note that $\tilde \Delta$ is a Hermitian  matrix.  Obviously, such an  approach   involves the invertibility   of the  matrix $\Delta$   thus we can consider the subsystem  $\mB$ provided $N-n\leq n$, in the other case we should   take  the   subsystem $\mA$  and use  the fact that  for the  pure  state the   spectrum of  $\rho_\mB$   is equivalent   (up to irrelevant zeros) to the  spectrum of $\rho_\mA$, see also  the discussion  in  \cite{b15b}; later on, using  a different  approach, we will  avoid this  technical problem. Now,  from  \eqref{ee20}  we can    immediately find  the R\'enyi entropies  
\be
\label{eee7}
R_\alpha(t)=\sum_{j=n+1}^N R_{\alpha}[\xi_j(t)],
\ee 
where   $R_{\alpha}[\xi_j(t)]$  has  the form as in \eqref{e84}. 
\par 
 In consequence,    the modular entropy as  well as the  capacity read
 \be
 \label{ee7}
 S_\alpha(t)=\sum_{j=n+1}^N S_{\alpha}[\xi_j(t)], \quad 		 C_\alpha(t)=\sum_{j=n+1}^N C_{\alpha}[\xi_j(t)],
 \ee
where $S_{\alpha}[\xi_j(t)], C_{\alpha}[\xi_j(t)]$  are of the form  \eqref{ee5} and \eqref{ee6}  (i.e. with the replacement    $\xi(t) \rightarrow \xi_j(t) $).   In  particular,     we immediately  obtain  the  formula for entanglement fluctuation  $C(t)\equiv C_1(t)$ of the system.
\par
In what follows  we will use another approach   based on the correlation matrix and symplectic  spectrum; it avoids   a direct computation  of the  spectrum    and   has been successfully  applied in  the study of  the entanglement entropy  and abrupt quenches, see e.g.  \cite{b12a0,b12aa,b12f}.     First, we define the time-dependent  correlations (covariance)  matrix 
\be
\Gamma=
\left(
\begin{array}{cc}
Q&R\\
R^T&P
\end{array}
\right),
\ee 
where  $Q_{ij}=\langle x^ix^j \rangle $, $P_{ij}=\langle p_ip_j\rangle $, $R_{ij}=1/2\langle \{x^i,p_j\}\rangle  $  are real and symmetric $N\times N$  matrices.  Using the results of Sec. \ref{s2} we obtain, after straightforward computations, that 
\be
\label{e26}
Q=U^T\tilde QU,\quad P=U^T\tilde PU, \quad R=U^T\tilde RU;
\ee
  where  $U$ is  defined as in \eqref{ee19} and   $\tilde Q,\tilde P,\tilde R$ are diagonal matrices with the following  diagonal elements 
\be
\label{e28}
\tilde Q_{kk}=\frac{b_k^2}{2\sqrt{\lambda_k(t_0)}},\quad \tilde P_{kk}=\frac1 2 \left(\frac{\sqrt{\lambda_k(t_0)}}{b_k^2}+\frac{\dot b_k^2}{\sqrt{\lambda_k(t_0)}}\right), \quad\tilde  R_{kk}=\frac{b_k\dot b_k}{2\sqrt{\lambda_k(t_0)}},
\ee
 for $k=1,\ldots,N$.  
 Since the matrix $\Theta$ is skew-symmetric the  covariance  matrix  of the reduced density   of the system $\mathcal B$ is  given by   the restriction  of the initial one   to the subsystem $\mathcal {B}$   
 \be
\Gamma_\mB=
\left(
\begin{array}{cc}
Q_\mB&R_\mB\\
R^T_\mB&P_\mB
\end{array}
\right),
\ee 
where $(Q_\mB)_{kl}=Q_{kl}$,  $(R_\mB)_{kl}=R_{kl}$ and  $(P_\mB)_{kl}=P_{kl}$ for $k,l=n+1,\ldots, N$. Next,  following the procedure   based on the symplectic transformation, see e.g. Refs. \cite{b12aa,b12f,b15bb}, we construct the matrix  $\tilde \Gamma_\mB=iJ\Gamma_\mB$,   where $J$ is given by  
 \be
J=
\left(
\begin{array}{cc}
0&I\\
-I&0
\end{array}
\right),
\ee 
and $I$ is  $(N-n)\times (N-n)$ identity matrix.     Now,    the spectrum of the matrix $\tilde \Gamma_\mB$ consists of elements $\pm\gamma_k$,  $k=n+1,\ldots,N$ and the R\'enyi  entropies take the form
\be
\label{ee16}
R_\alpha=\frac{1}{\alpha -1}\sum_{k=n+1}^N\ln\left((\gamma_k+1/2)^\alpha-(\gamma_k-1/2)^\alpha \right).
\ee
By means of \eqref{ee16}  we can  easily find the modular entropy and capacity; for example,   the entanglement fluctuation   reads
\be
\label{ee17}
C\equiv C_1= \sum_{k=n+1}^N(\gamma_k^2-\frac 14)(\ln(\gamma_k+1/2)-\ln(\gamma_k-1/2))^2.
\ee
In view of this and eqs. \eqref{e26}, \eqref{e28}    the temporal evolution of the modular entropy and capacity  is algebraically determined by  the solutions (and their derivatives) of  the EMP equations. Moreover, 
the quantities obtained in this way   coincide with the formulae \eqref{eee7} and \eqref{ee7}   after the identification 
\be
2\gamma_k=\frac{1+\xi_k}{1-\xi_k}.
\ee
Finally,  let us note that  $0 \leq C(t)\leq N-n$.    Moreover,  the state is pure  thus  $0 \leq C(t)\leq n$. In consequence,  for  the initial ground state   the capacity remains  bounded
\be
\label{e25}
 0 \leq C(t)\leq \min (n, N-n).
 \ee
\par 
 Concluding, eqs. \eqref{ee5} for $N=2$  and \eqref{ee7}  (equivalently \eqref{ee16} and \eqref{ee17})   for higher $N$    enable  us to analyse   the  evolution more explicitly  provided  we have  solutions  $b_j(t)$ to the EMP equation. Such a situation holds for some special forms of  frequencies.  In what follows   we will discus  both   abrupt   and continuous  examples and  concentrate on    the entanglement fluctuation  and its relation to  the  vN entropy; this  completes  the results for the finite-dimensional spaces and fit into  the field theory problems.
\subsection{ Frequency and coupling jumps}
\label{s3c}
Although the general dynamics  of the entropy and capacity  in the quench phenomena is  complicated, some qualitative description  can be obtained if  we know  the explicit form of the solutions to the EMP equations.  To see this let us consider the  abrupt change.    Namely, let us analyse this evolution  for  the model \eqref{ee8}   with  a  sudden quench, i.e. where  $\omega(t),k(t) $  change, at time $t_0=0$,   from constant  values $(\omega(i),k(i)))$ to  another constant values  $(\omega(f),k(f))$.     For the abrupt quench  the solutions of the EMP equations  \eqref{ee9}  with the initial conditions \eqref{e15}   read
\be
\label{e16}
b_j(t)=\sqrt{r_j\cos(2t\sqrt{\lambda_j(f)})+s_j},
\ee  
 where $\lambda_j(i), \lambda_j(f)$   are the eigenvalues of $\Lambda$  before and after quench and $r_j=(\lambda_j(f)-\lambda_j(i))/(2\lambda_j(f))$,    $s_j=(\lambda_j(f)+\lambda_j(i))/(2\lambda_j(f))$.   
 \par
Both the entanglement and capacity  are functions of $b(t)$'s   and their derivatives. Thus, due to the formula \eqref{e16},  we expect some distinguished periodicities in the temporal evolution.
\par
\begin{figure}[!ht]
\begin{center}
\includegraphics[width=0.95\columnwidth]{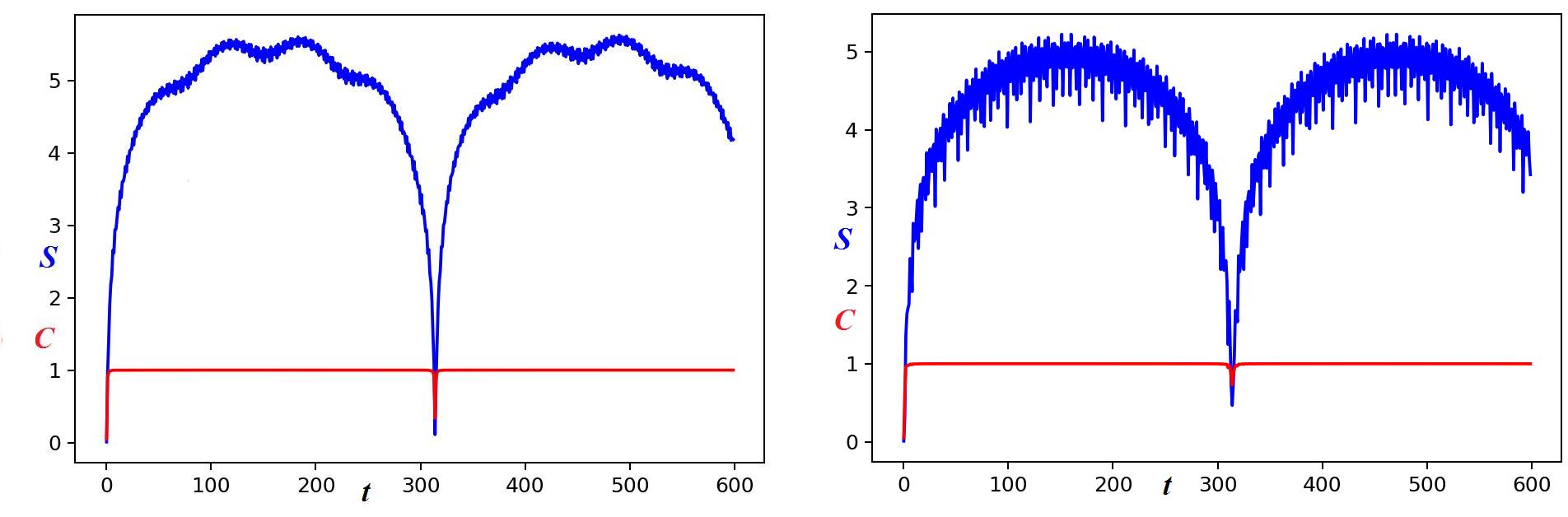} 
\end{center}
\caption{\small{Abrupt  quench - temporal evolution: $N=4, n=1$  and PBC ($\omega(i)=3$,  $\omega(f)=0.01$ ). Lines:    blue - entropy,  red - capacity.  The left panel $k(f)=10$. The right panel $k(f)=0.5$ }
\label{f1}
}
\end{figure}

  For the PBC the last frequency     $\lambda_N(f)=\omega^2(f)$ is distinguished. Indeed, it tends to zero as $\omega(f) $ is small, in contrast to   the previous  frequency   $\lambda_{N-1}(f)=\omega^2(f)+4k(f)\sin^2((N-1)\pi/N)$   when  $k(f)\gg \omega^2(f)$.      Under these assumptions $r_N,s_N$ tend to infinity  as $\omega(f)$ is close to  zero; thus  due to  eq. \eqref{e16}    we expect the  distinguished  periodicity $\mT=\pi/\omega(f)$ (in particular, it does not depend on $k(f)$).  This situation is presented in Fig. \ref{f1},    where $N=4$ and  $\omega(f)=0.01$ (i.e. $\mT\simeq  314$). Let us note that for the capacity (which is  bounded)  this  is less evident, especially, as  we pointed  out above,  for  small $k(f)$,  see the right panel in Fig. \ref{f1}. Finally,   it is worth to notice that for    $N\rightarrow \infty$    the situation changes: $\lambda_N$   is not distinguished,  and the above  reasoning breaks down; we return to this issue in the next sections.   
\par
For the DBC   the situation is quite different. Namely,   for $\omega(f)=0$ the $\lambda$'s remain non-zero.  However,   for  large $N$  the frequency  $\lambda_N $ tends to zero  implying that   the distinguished periodicity  should be of the form 
\be
\label{e27}
 \mT=\frac{\pi}{ 2\sqrt{k(f)}\cos(\pi N/(2(N+1)))}, 
\ee
 thus  it depends on  $k(f)$ (and does  not depend on  $k(i)$; in particular,  we can  put   $k(i)=k(f)=k$). For example,   for $N=100$  and   $k(f)=1$ we have  $\mT\simeq  100$ while   for $k(f)=9$  we obtain      $\mT\simeq 33$;    see Fig. \ref{f4} and \ref{f5}  in the Sec. \ref{s4bb}.  In general,   we have   $ \mT\simeq \frac{N}{\sqrt{k}}$  for large $N$,  thus  for  $k=1$ (lattice approximation)  we have $ \mT\simeq N$, see the next section.  

\section{Fluctuation of entanglement and field theory } 
\label{s4}
In this section  we  consider larger  values of  $N$.  This is interesting due to  the lattice  approximation to  quantum fields,  e.g.   a time-dependent  bosonic scalar field $\phi$.  Namely,  we discretize the field Hamiltonian, in general with   time-dependent mass $m(t)$,
\be
H=\frac{1}{2}\int dx (\pi^2+(\nabla \phi)^2+m^2(t)\phi^2),
\ee
   into a chain of harmonic oscillators by imposing a UV  cutoff $\epsilon$  and IR cutoff $N\epsilon$.  Then we  obtain a model  described by the Hamiltonian \eqref{ee8} with $k=1$ and $\omega^2(t)=m^2(t)\epsilon^2$.  To complete the  discretization procedure we need  boundary conditions; we take the  PBC for the  spacetime cylinder; the other choices are the DBC or NBC.    
\par 
\subsection{Constant frequency}
\label{s4a}
 For start  we consider   the constant mass,    which can be realized as a special case of the abrupt quench with  $\omega(i)=\omega(f)=\omega=const$. We will focus on more subtle   case when $\omega\rightarrow 0 $  which is related   to the 2-dimensional  CFT.   In this case the PBC  (or  NBC)   leads to   zero modes   which  cause the   divergence of the  entanglement entropy. For the DBC   there are no zero modes and  no infrared divergences arise when setting $\omega$  equals  zero (the divergence is in the limit $N\rightarrow \infty$ only). For   the entanglement capacity the situation is quite different.  Namely, as we noted   above the entanglement capacity is bounded thus  for  $\omega=0$  we expect a finite value. 
\par
Using  methods of the  2-dimensional  CFT it has been shown \cite{b9h} that at  the leading order  of the UV cutoff  parameter $\epsilon$ the capacity and entropy are equal to each other. Namely,     for the PBC  (circle of the length $L$)  we have
\be
\label{e10}
S=\frac 13 \ln\left(\frac {L}{\pi \epsilon} \sin\left (\frac{\pi l}{L}\right)\right)+O(1) =C.
\ee
In particular,  for the infinite length  $L\rightarrow \infty$  we obtain $S=\frac 13 \ln(l/\epsilon) +O(1)=C$;  such a behavior  has been confirmed in Ref.  \cite{b9h}  by considering  the infinite harmonic chain. Here, we  analyse this problem for $L$ finite: $L=N\epsilon$ (and $l=n\epsilon$). 
For the  DBC    instead of  eq. \eqref{e10} we expect the relation 
$S=\frac 16 \ln\left(\frac {L}{\pi \epsilon} \sin\left(\frac{\pi l}{L}\right)\right)+O(1) =C$. In the former case we take $\omega=0.001$ and next  analyse the limit as $\omega\rightarrow 0$, in the Dirichlet  case we can put directly $\omega=0$ from  very beginning.   
\par
\begin{figure}[!ht]
\begin{center}
\includegraphics[width=0.95\columnwidth]{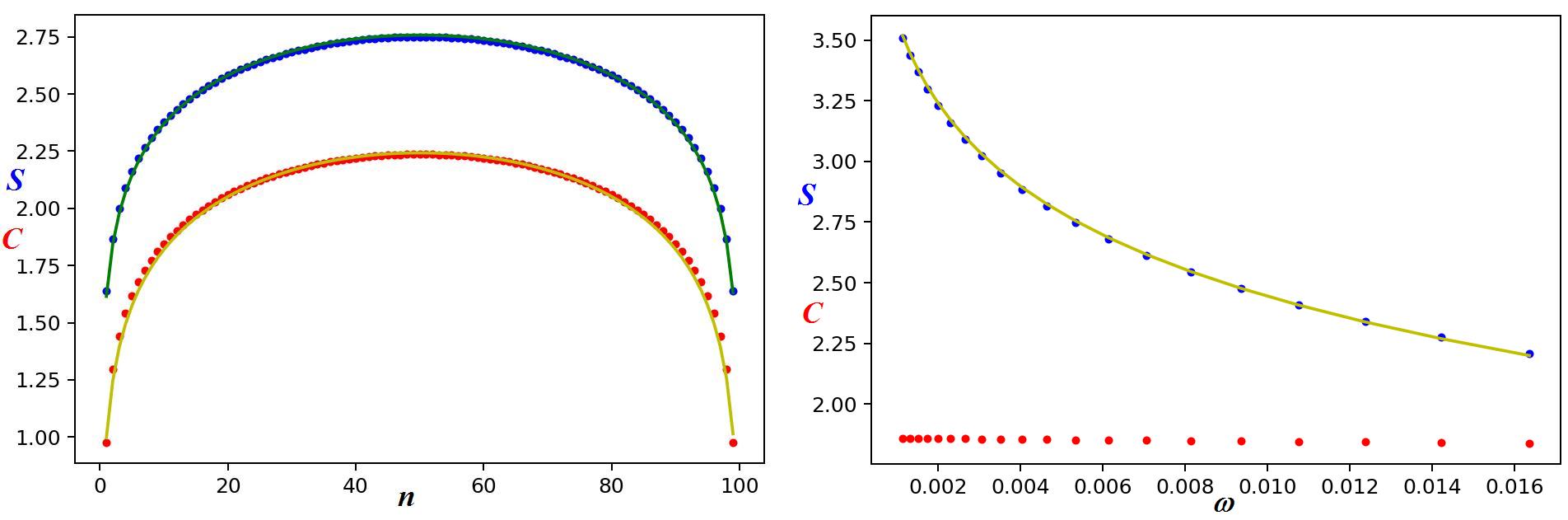}
\end{center}
\caption{\small{$N=100$  and PBC ($\omega=0.001$ and $k=1$).  The left panel:  entropy  - blue data points    (green line corresponds to \eqref{e10}), capacity  - red data points (yellow line corresponds to \eqref{e10}). The right    panel:    omega dependence, entropy -   blue date points, 	capacity - red date points.}
\label{f2}
}
\end{figure}

The results for the  PBC  are  presented  in the left panel of   Fig. \ref{f2} (for $N=100$). 
We see that  the   entropy and capacity   coincide  with eqs. \eqref{e10} (represented here  by green  and yellow lines, respectively). For the small ratio  $n/N=l/L$ this formula   exhibits   a logarithmic scaling,  i.e. it diverges, in contrast to the area law.   Moreover,  for the PBC  the entropy increases (for fixed $n$  and  $N$) as $\omega$ tends to zero   as is presented in the right panel of  Fig.  \ref{f2},   in contrast to the capacity which  is bounded (cf.  eq. \eqref{e25}).
\par  For the DBC  we  can put directly  $\omega=0$,  then  both entropy and capacity are finite and  their  behavior  agrees with the  above theoretical prediction (i.e. eq. \eqref{e10} with $1/3$ replaced  by $1/6$). 
\subsection{Quenches}
\label{s4b}
Now, let us pass to  the case  of   oscillatory chains with time-dependent $\omega(t)$. Such  models can  approximate  quench protocols in the   field theory  where the mass varies in time or in expanding spacetimes.    We will focus on the more interesting case where  the final theory is massless. Due to zero modes we will consider the  DBC  and PBC separately; in the latter  case,    the final frequency will be small but not equal to zero. 

\subsubsection{Abrupt  quench }
\label{s4bb}
First, we consider the abrupt  quenches; namely,   we start with some   $\omega(i)$  and  at time $t_0=0 $  then there is a sudden change of the frequency  to a small or zero  value.   Such a situation is motivated by the CFT methods  developed in the  study of the temporal evolution of the R\'enyi entropies. For the    global quench and  the subsystem which is an  interval on the   infinite line    the  field  approach yields  \cite{b9h}
 \be
 \label{e24}
S= \textrm{const}+ \left\{
\begin{array}{cc} a_St \quad t<t^*, \\
b_Sn\quad  t>t^*;
\end{array}
\right. \quad 
C= \textrm{const}+ \left\{
\begin{array}{cc} a_Ct \quad t<t^*,\\
b_Cn\quad  t>t^*;
\end{array}
\right.
\ee 
with  some constants  $a_S,b_S,a_C,b_C$ and $t^*=n/2$;  moreover, the relation   $a_S=a_C$ holds. However,   the analysis performed  in \cite{b9h} suggests that  the  situation is more subtle  and the slopes  $a_C$ and  $a_S$ are  not equal  (and, consequently, the saturation values too).  Here, using the results presented in Sec.  \ref{s3}, we  analyse this  problem in the finite case and various  boundary conditions; in particular,  we obtain some theoretical predictions for these constants.  
\par
\begin{figure}[!ht]
\begin{center}
\includegraphics[width=0.95\columnwidth]{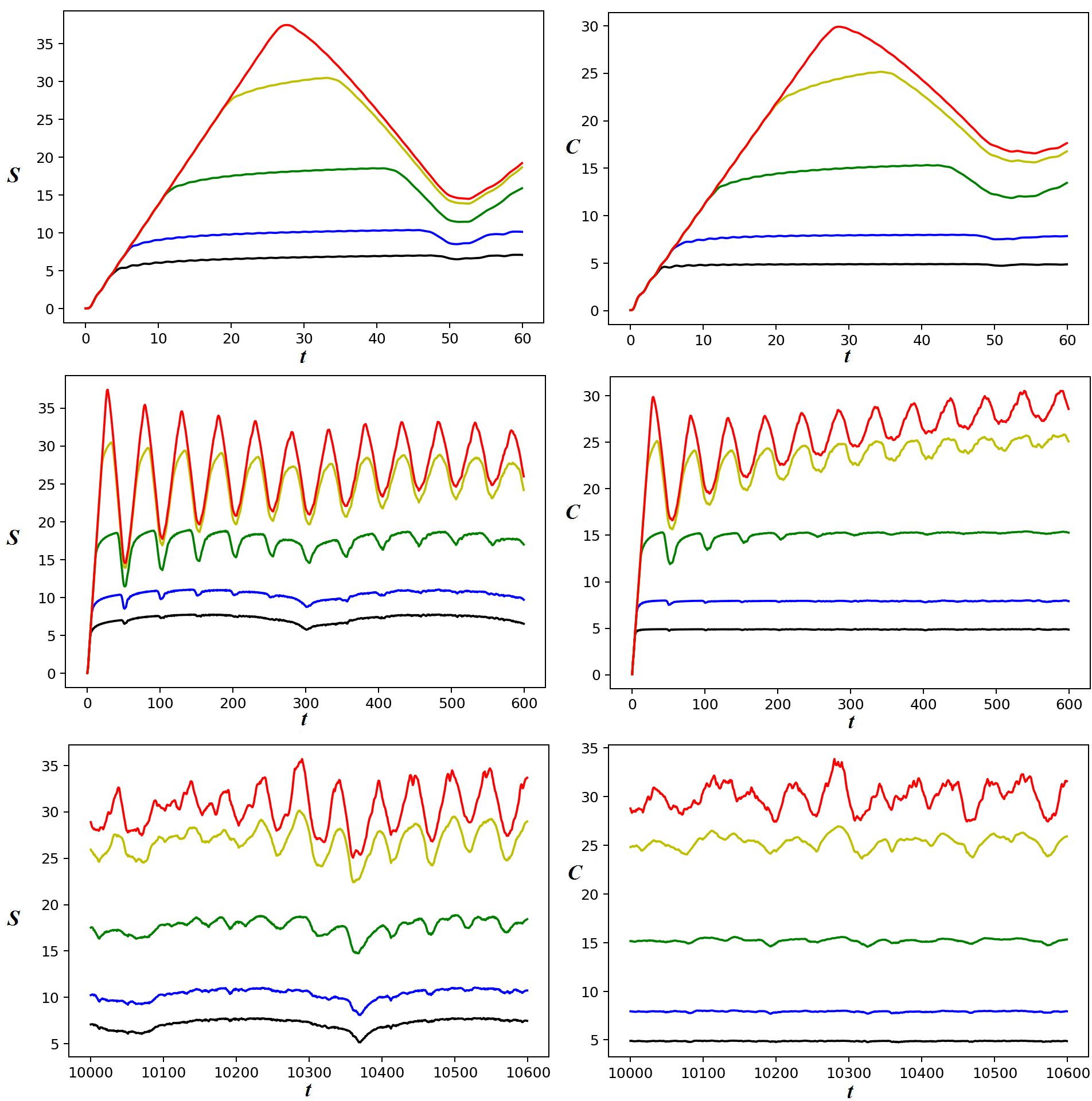}
\end{center}
\caption{\small{Abrupt  quench - temporal evolution: $N=100$  and PBC ($\omega(i)=3$, $\omega(f)=0.01$ and $k=1$). Lines:  black  $n=6$ ,  blue  $n=10$,  green $n=20$,  yellow  $n=36$, red    $n=50$.  The left panels  entropy. The right panels  capacity. }
}
\label{f3}
\end{figure}

We start  with  temporal evolution of the entropy and capacity   for various $n$  and PBC. In Fig. \ref{f3} (top panels),  we see that there is indeed  the linear growth of the  both  quantities (more precisely, just after  the quench there is parabolic-like growth); however,   as in Ref. \cite{b9h},  we observe  that the slopes $a_S$ and $a_C$   are, in general,  not equal; it   depends on the initial  frequency.    Approximately, after    time $t>n/2$     and for   lower  $n$  there is a  logarithmic-type   growth  instead of  the constant value  (both for the entropy and capacity);   this can be  related to zero modes appearing for the PBC,    see e.g. the discussion in \cite{b12aa} and references therein. Next,  both quantities oscillate; after longer time   these oscillations are around   saturation  values (bottom panels in Fig. \ref{f3}).  Note that the values of  both  the entropy  and capacity  increase with  $n$ ($l$); namely,  for $n<N/2$ they are less  than   for  $n=N/2$ (for $n>N/2$ they coincide with the ones for $N-n$). Finally, let us note that the  dynamics of capacity is bounded as   in eq. \eqref{e25}. 
\par
\begin{figure}[!ht]
\begin{center}
\includegraphics[width=0.95\columnwidth]{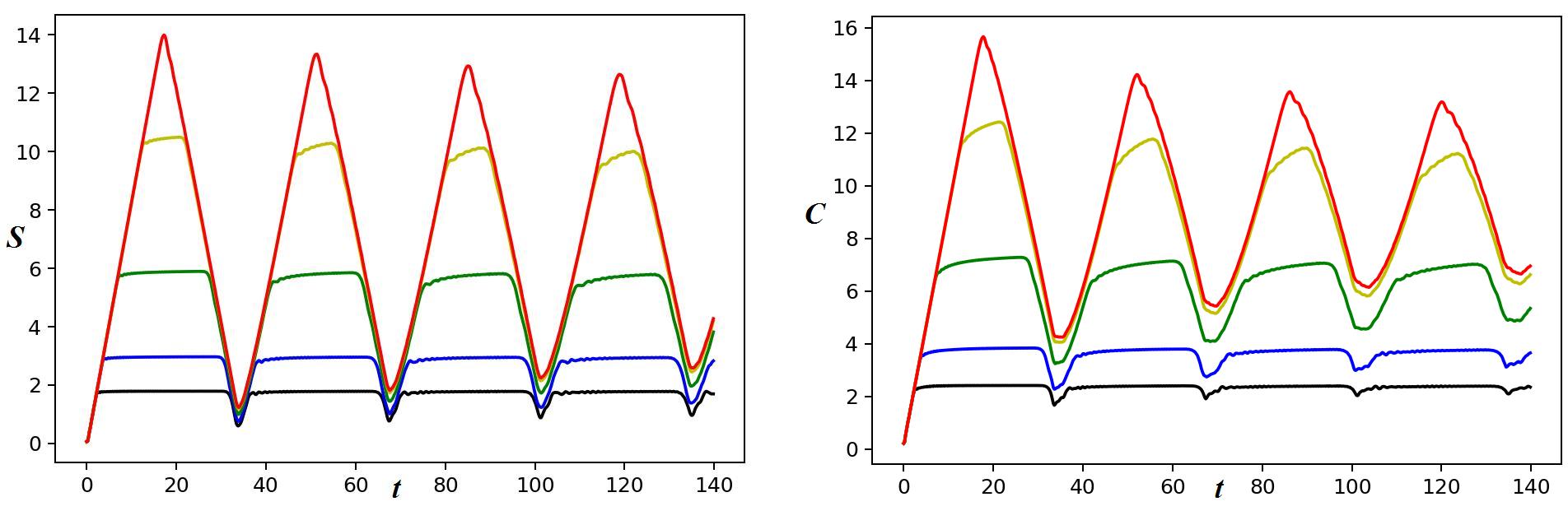} \hspace{0.5cm} 
\\
\end{center}
\caption{\small{Abrupt  quench - temporal evolution: $N=100$  and DBC ($\omega(i)=3$, $\omega(f)=0$ and $k=9$). Lines:  black  $n=6$ ,  blue  $n=10$,  green $n=20$,  yellow  $n=36$, red    $n=50$.  The left panels  entropy. The right panels  capacity.  }
\label{f4}
}
\end{figure}

\begin{figure}[!ht]
\begin{center}
\includegraphics[width=0.95\columnwidth]{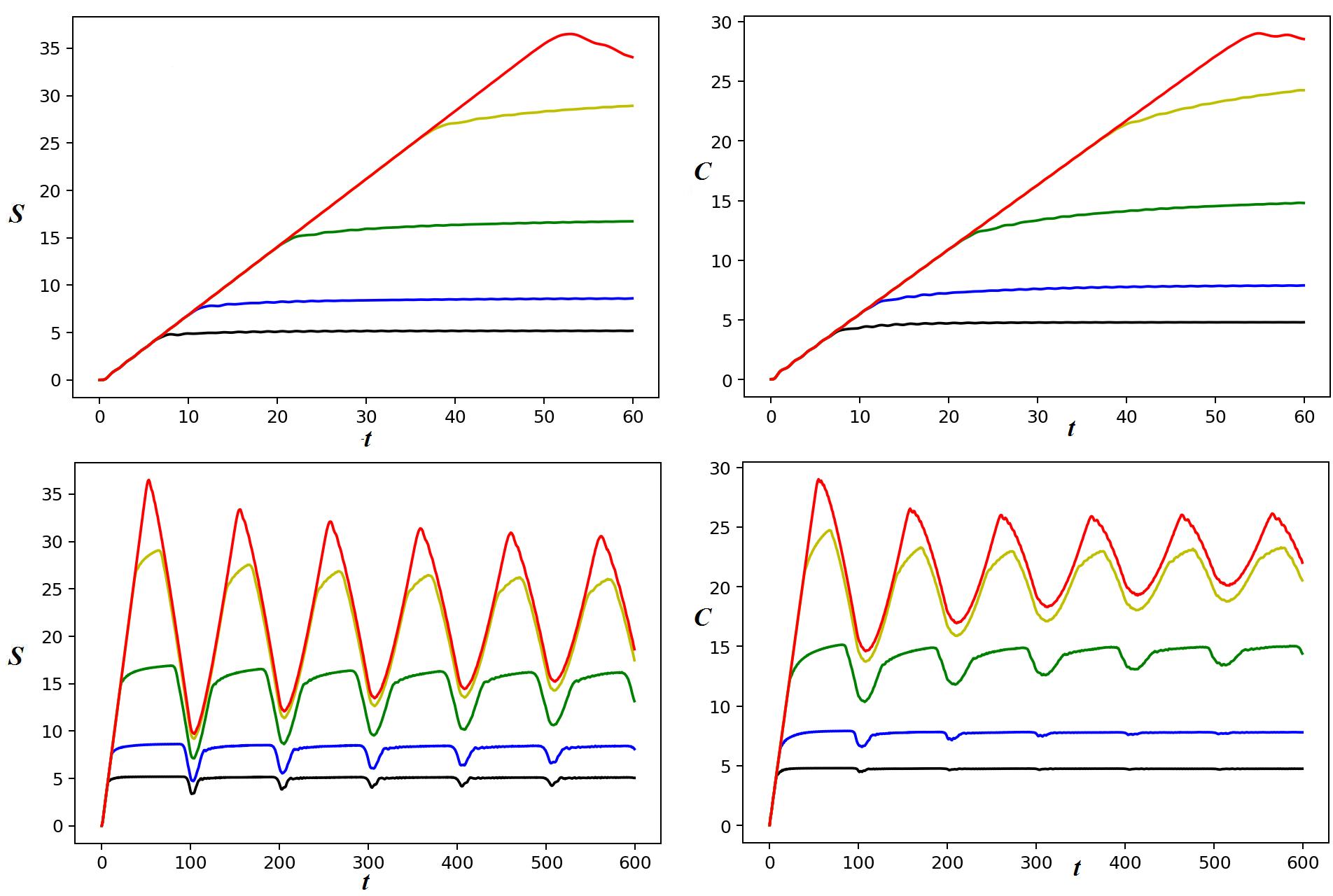}
\end{center}
\caption{\small{Abrupt  quench - temporal evolution: $N=100$  and DBC ($\omega(i)=3$, $\omega(f)=0$ and $k=1$). Lines:  black  $n=6$ ,  blue  $n=10$,  green $n=20$,  yellow  $n=36$, red    $n=50$.  The left panels  entropy. The right panels  capacity. }
\label{f5}
}
\end{figure}

 
 Let us compare these  results with  the ones for  the DBC. In this case   we can put the final frequency zero $\omega(f)=0$ (the entropy is finite).  Then,  in agreement with the discussion presented in Sec. \ref{s3c},    we observe the distinguished periodicity $\mT\simeq  N/\sqrt{k}$  for large $N$; namely,  for $k=1$ ($N=100$)  we have  $\mT\simeq 100$ , see Fig.  \ref{f5}, while  for $k=9$ it gives  $\mT\simeq 33$, see Fig. \ref{f4}. This special periodicity  is related to   local minima,  such that  the  quaintness coincide for all $n$  (especially when the initial frequency is not too  large, see see Fig. \ref{f4}),  thus  at these points we can  observe  the area law ($n^0$). In contrast to this,   at  local maxima  or for  further times   the  area law is  broken.  Moreover,    the initial linear growth is up to the first maximum which, see Sec. \ref{s3c}, depends on $k$  (in contrast to  the  PBC case), it is  around the half of the period $\mT$, i.e.  $t=N/2$ (in agreement with  the  maximal linear growth resulting  from the field approach, i.e.  $t=n=N/2$).   For   small  $n$  the dynamics exhibits plateaus (instead of maxima)  and finally oscillates  around  saturation values. 
 \par 
\begin{figure}[!ht]
\begin{center}
\includegraphics[width=0.95\columnwidth]{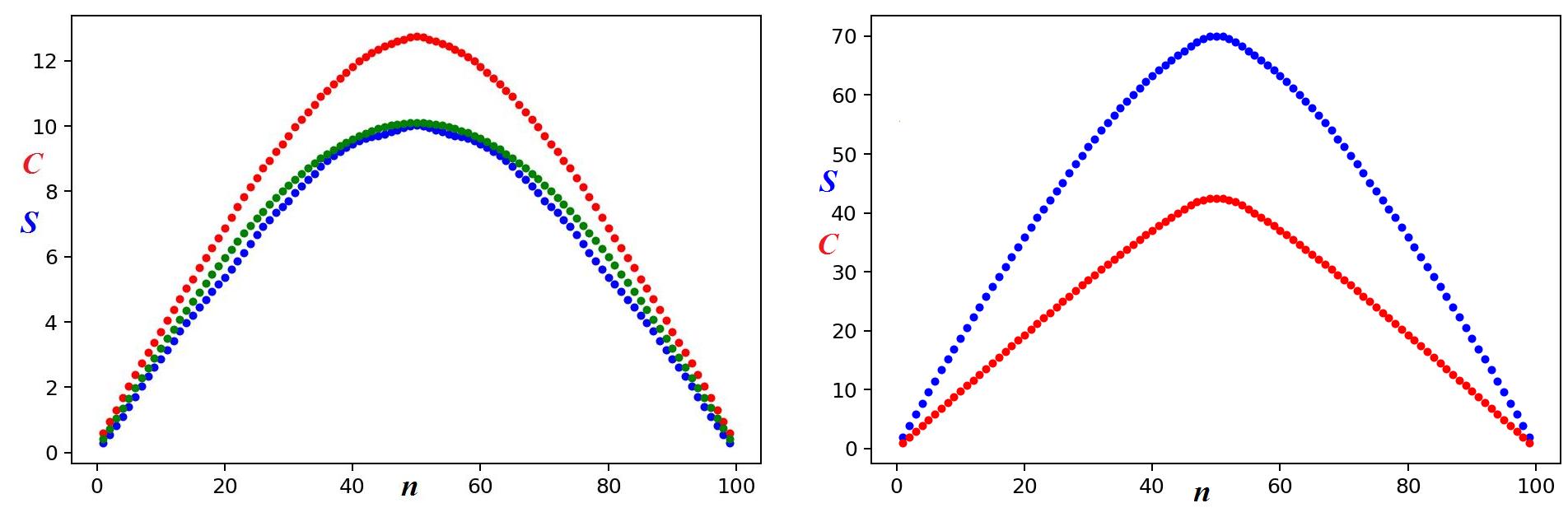} 
\\
\end{center}
\caption{\small{Abrupt  quench -  time slices,  $N=100$     and DBC ($k=1$, $\omega(f)=0$) and $t=10150$.  Date points:   blue - entropy, red  - capacity. Left panel $\omega(i)=1$ and     green  date points:  $10\sin(\pi n /100)+0.1$. The right panel  $\omega(i)=10$. } 
\label{f6}
}
\end{figure}

\begin{figure}[!ht]
\begin{center}
\includegraphics[width=0.95\columnwidth]{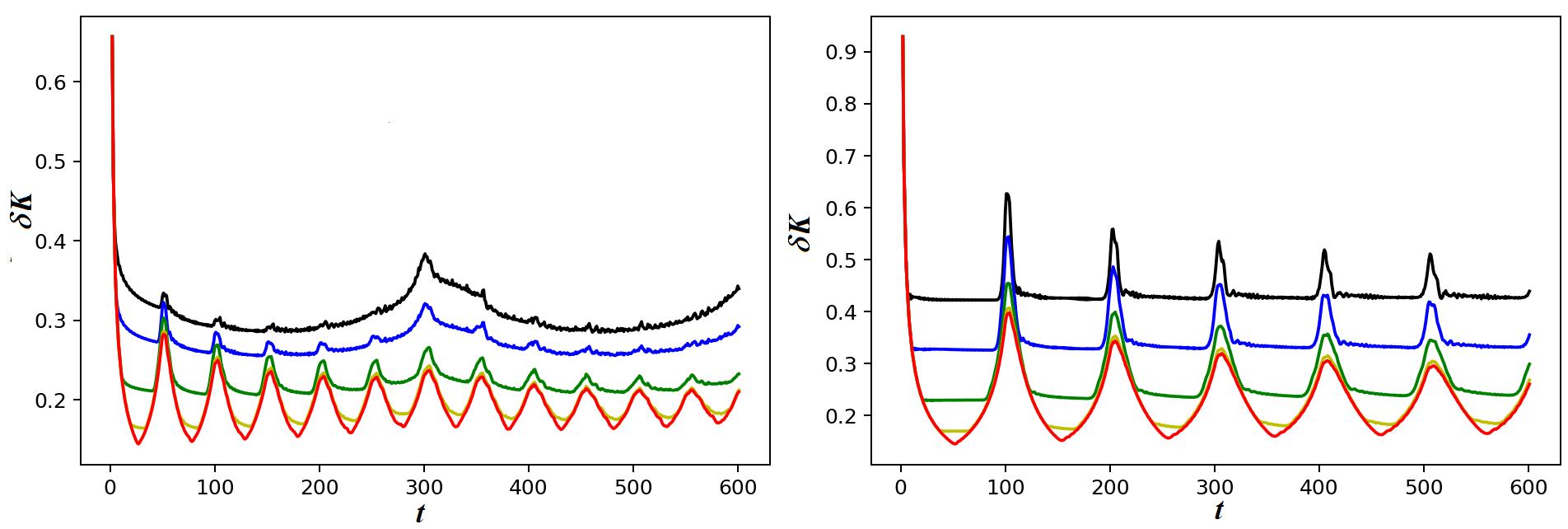} 
\end{center}
\caption{\small{Abrupt  quench - temporal evolution of relative fluctuations, $N=100$ ($k=1$). Lines:  black  $n=6$ ,  blue  $n=10$,  green $n=20$,  yellow  $n=36$, red    $n=50$.  The left  panel: PBC ($\omega(i)=3$, $\omega(f)=0.01$). The right panel: DBC ($\omega(i)=3$, $\omega(f)=0$). }
\label{f25}
}
\end{figure}

 These results  can be confirmed  by  considering time-fixed slices, see Fig. \ref{f6}. Then  just    after the quench the  behavior of the entropy and capacity  resembles the one  for the finite mass, i.e. it is constant   w.r.t.  $n$;  the area law holds.  For larger time   the entropy (capacity)  behaves linearly for  more   $n$; however,  around $n=N/2$  this behavior changes. This  is related to the finite size of the whole system; instead of the  linear growth  rather $\sin(\pi n/N)$  appears (see green data points in Fig.  \ref{f6}). Moreover, the symmetry $n\rightarrow N-n$ is   preserved  for  entropy and capacity. In consequence, the entropy and capacity      of the subsystems   $\mA$ and $\mB$ coincide (the state is pure). 
Finally in order to compare  better the dynamics of the entropy and capacity    we   plot  the magnitude of  relative quantum entanglement fluctuations $\delta K=\sqrt C /S$, see Sec. \ref{s1}. In Fig.  \ref{f25} we see that after   some  initial time  the relative entanglement fluctuations are  small and exhibit some revival time; moreover, they are larger for    lower $n$  (for the DBC the maxima are related to the   distinguished periodicities discussed above).
\par
\begin{figure}[!ht]
\begin{center}
\includegraphics[width=0.95\columnwidth]{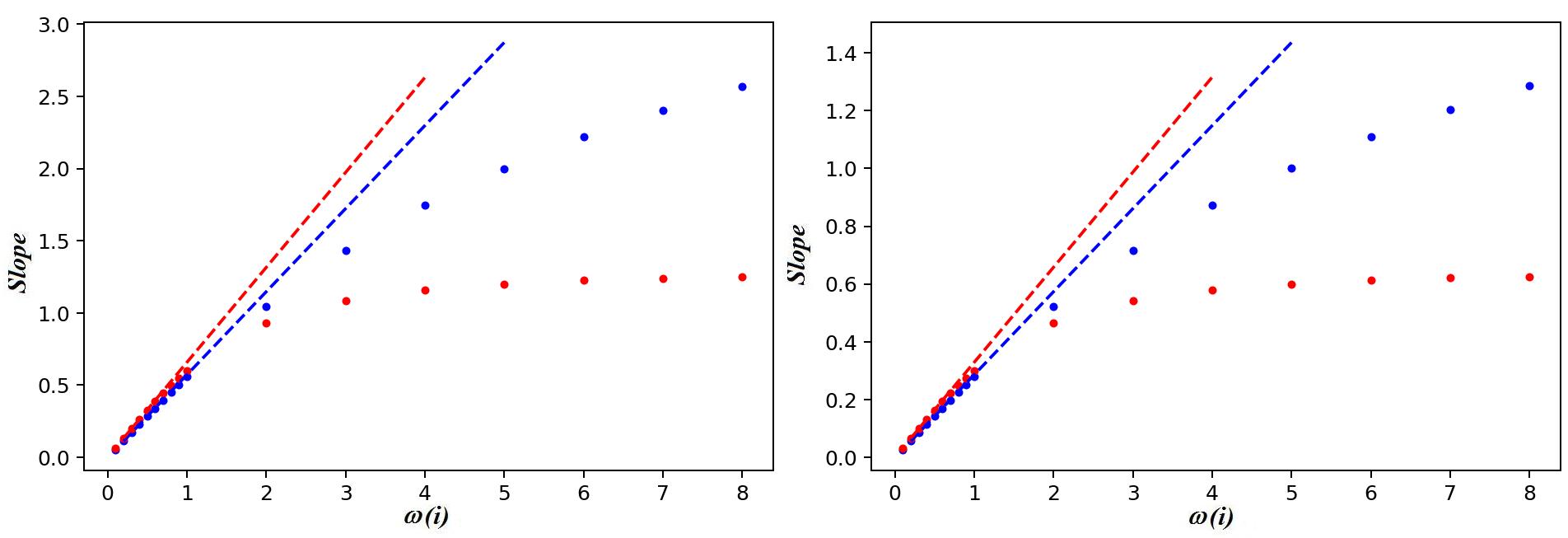} 
\end{center}
\caption{\small{The initial slope for the abrupt  quench,    $N=100$.   Lines:   blue - entropy, red  - capacity. Left panel PBC ($\omega(f)=0.01$);  right panel DBC  ($\omega(f)=0$).} 
\label{f23}
}
\end{figure}

 
Now, let us return  to the   field theory picture.   Based on the quasiparticles  considerations  we expect  the relation  \eqref{e24}  to hold (at least for small $n$).     To this end    let us compute the  initial slope  of $S(t)-S(0)$ as well as $C(t)-C(0)$  w.r.t. $\omega(i)\geq 0.1$   for  the   DBC  and  PBC (in the latter case  $\omega(f)=0.01$ is fixed). The results are presented in Fig. \ref{f23}.  First, we see that  the slope  coefficient behaves linearly for   $0.1\leq \omega (i)<1$  and it is smaller   for the entropy than for the capacity.   However,  for the  capacity which is slowly increasing,  the linearity is broken earlier; thus to analyse it we restrict ourselves  to the interval    $0.1\leq \omega (i)\leq 0.5$.   More preciously, the dashed lines in Fig. \ref{f23} for the  PBC are given by $0.575\omega(i)+0.003$ and $0.657\omega(i)+0.0004$ for  the entropy  and capacity, respectively;  while for the DBC we have $0.287\omega(i) +0.001$ and $0.329\omega(i) +0.0004$, respectively.  In view of this  the  slope $a_S=0.575$  for the  entropy  and PBC is   two times greater than for the DBC ($2a_S=0.574$) and agrees with theoretical prediction  $a_S^t=(\pi-2)/2\simeq 0.571$ presented in \cite{b12aa}; for the capacity we have  also the same  relation  ($0.657$   and   $2a_C=0.658$). Even more in the non-linear region the slope for the  PBC   is   twice  the one for the DBC (both for the entropy and capacity), cf. the  left and right panels in Fig. \ref{f23}.   Next,    as $\omega(i)$ becomes greater  than one the linearity is more  and more  broken   and   both slopes  meet; then the  entropy increases while capacity stabilizes. In consequence, for   large $\omega(i)$ the entropy  slope is greater than for the  capacity. 
\par
\begin{figure}[!ht]
\begin{center}
\includegraphics[width=0.95\columnwidth]{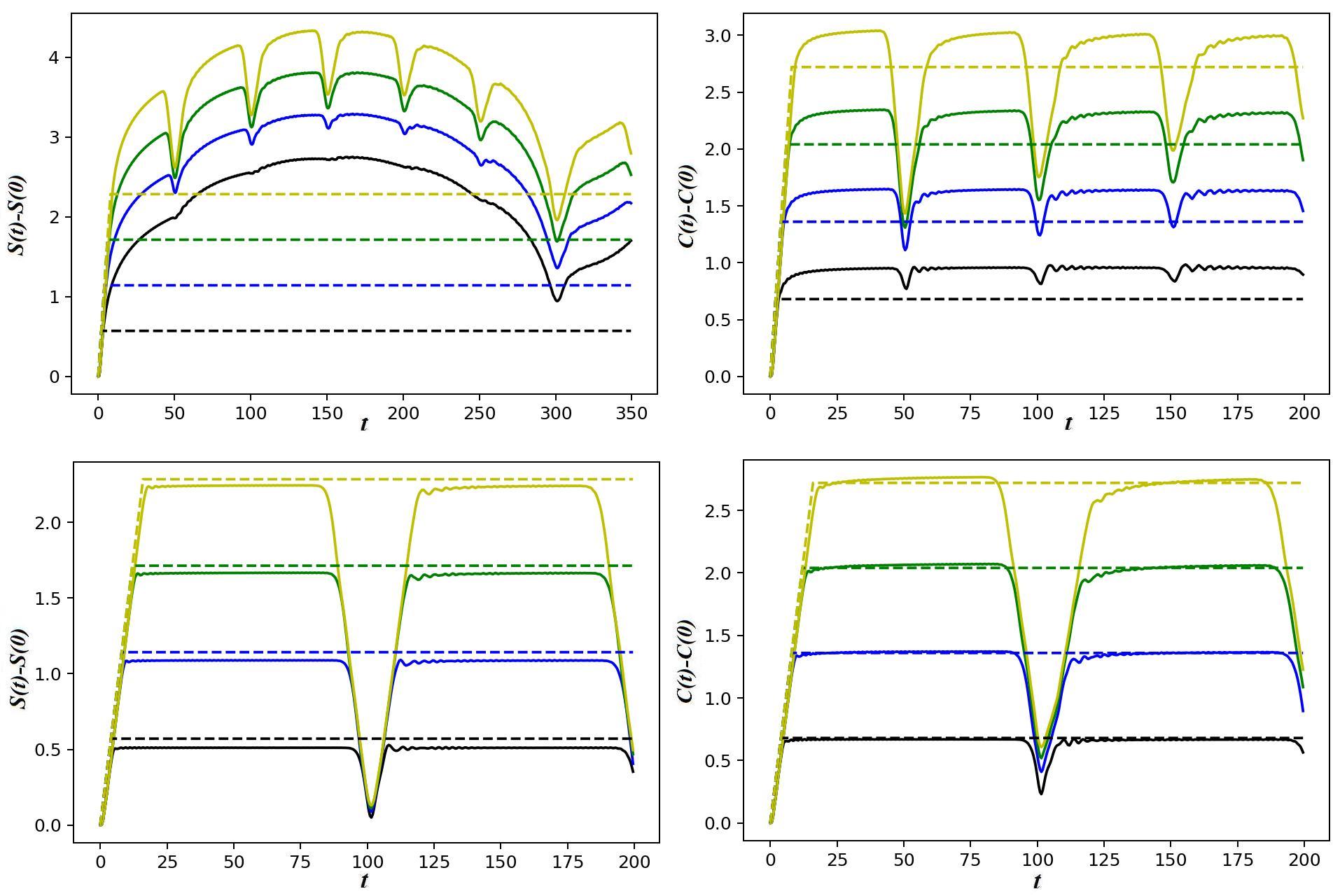} 
\end{center}
\caption{\small{Abrupt  quench - temporal evolution: $N=100$  ($\omega(i)=0.5$ and $k=1$).  The top panels PBC ($\omega(f)=0.01$). The  bottom  panels  DBC ($\omega(f)=0$).  The left panels  entropy. The right panels  capacity.  Lines:  black  $n=4$ ,  blue  $n=8$,  green $n=12$,  yellow  $n=16$. Dashed lines  correspond to eq.  \eqref{e24} with the theoretical parameters $a^t_S,a^t_C$ . }
 \label{f24}
}
\end{figure}

 
Let us return to the slope   of the  capacity.   Using  the considerations  from  Ref. \cite{b12aa} (basing on the  quasiparticles approach \cite{b15c})   and formula \eqref{ee3} we find   the  initial  slope  for  the capacity and the PBC: 
\be
a_C^t=\partial_\alpha^2(\alpha-\cot(\pi/4\alpha))(1)=\pi(1-\frac{\pi}{4})\simeq 0.674,
\ee
thus  it   quite well  agrees with the above  numerical result $a_C=0.657$ obtained by the lattice approximation.   It remains  to analyse the behavior  for further time;  then   for the PBC $b_S,b_C  $ we   should  have $b_S=a_S/2$ and $b_C=a_C/2$ while for  the DBC    $b_S=a_S$ and $b_C=a_C$.  In Fig. \ref{f24} we present  this situation  using $a_S^t$ and $a_C^t$ coefficients (for the PBC and DBC, respectively).  Let us stress that   these theoretical predictions  match quite well only for some  range of the initial frequencies, i.e. around $\omega(i)=1$ for the  entropy and $\omega(i)=0.5$ for the  capacity, see Fig. \ref{f24} for $\omega(i)=0.5$; such a restriction  is related to the validity of  lattice approximation ($\omega(i)$ should be not too small or large, see e.g. the discussion in Refs. \cite{b12a0,b12aa})   and the fact the quasiparticles model gives a  qualitative description.   A further investigation  of the field theoretical considerations    presented  recently in Refs.  \cite{b9h,b12h} can  give more insight into this problem. 
\par
Summarizing, for the abrupt quench  and small $n$ the  relation \eqref{e24}  approximates the dynamics of  the  entropy and capacity.   More precisely, we have a linear initial  growth and for further times the oscillations are about saturation values (up to  logarithmic  period  for the PBC);   however,    the initial slope coefficients are different for the entropy and capacity. For some range of initial frequencies  the capacity  slope can be also  obtained  by means of the quasiparticles picture.

\section{Continuous protocols}
\label{s5}
In this section we analyse the aforementioned  issues in the case of continuous protocols. To this end we consider two examples  which enable us  more  analytic considerations. 
 The first one with  a continuous frequency change at time $t_0=0$  such that it asymptotically tends to zero. 
In the second case we consider the situation when  at minus and plus infinities  the frequency is zero (the massless field).  For better transparency we describe  the   DBC  and we  mention  only  the necessary  changes for the PBC  or NBC.
\par
To   this end we put    $k=1$ and  take
\be
\label{e12}
\omega^2(t)=\frac{2}{\varepsilon^2\cosh^2(t/\varepsilon)},
\ee
in the Hamiltonian \eqref{ee8}. 
Then   $\omega(t)$ is a  bell shaped function with the maximum at $t=0$ and tending  to zero at infinities. In particular,  we obtain  the Dirac delta   limit
\be
\lim_{\varepsilon\rightarrow 0}\omega(t)=\pi\sqrt{2}\delta(t).
\ee
In order to obtain a continuous (differentiable)  quench  protocol  we  take 
 \be
 \label{ee10}
 \tilde \omega^2(t)= \left\{
\begin{array}{cc}
\omega^2(i)\equiv 2/\varepsilon^2 & \quad \textrm{for} \quad  t\leq 0,\\ 
\omega^2(t)& \quad \textrm{for} \quad  0< t.
\end{array}
\right.
\ee  
Then   the initial frequency  $\omega(i)$  is quenched to  zero  at infinity.   To analyse such a  model we need the   solutions of the EMP equations \eqref{ee9}  with
\be
\lambda_j(t)=\tilde\omega^2(t)+a_j,
\ee 
where,  for the DBC,  $a_j=4k\cos^2(\frac{j\pi}{2(N+1)})>0$, $j=1,\ldots,N$. 
  The corresponding  functions $\tilde b_j(t) $   can be   given  explicitly. Indeed,  by means  of the  results of  Ref.  \cite{ba}, we   obtain     
   \be
\tilde b_j(t)= \left\{
\begin{array}{cc}
1 & \quad \textrm{for} \quad  t\leq 0,\\ 
b_j(t)& \quad \textrm{for} \quad 0<t;
\end{array}
\right.
\ee
  where $b_j(t)$, for $j=1,\ldots,N$,  are of the form
\be
b_j^2(t)=\left(1+\frac{\tanh^2(t/\varepsilon )}{a_j^2\varepsilon ^2}\right)\left(1-\frac{\sin^2\left(a_jt+\tan^{-1}(\frac{\tanh(t/\varepsilon)}{a_j\varepsilon})\right)}{(1+a_j^2\varepsilon^2)^2}\right).
\ee 
Note that for the PBC we have  $a_N=0$   and  thus we have to modify   $b_N$ as follows
\be
b^2_N(t)=\left(1-\frac{t}{\varepsilon}\tanh(t/\varepsilon)\right)^2+2\tanh^2(t/\varepsilon).
\ee
\par
\begin{figure}[!ht]
\begin{center}
\includegraphics[width=0.95\columnwidth]{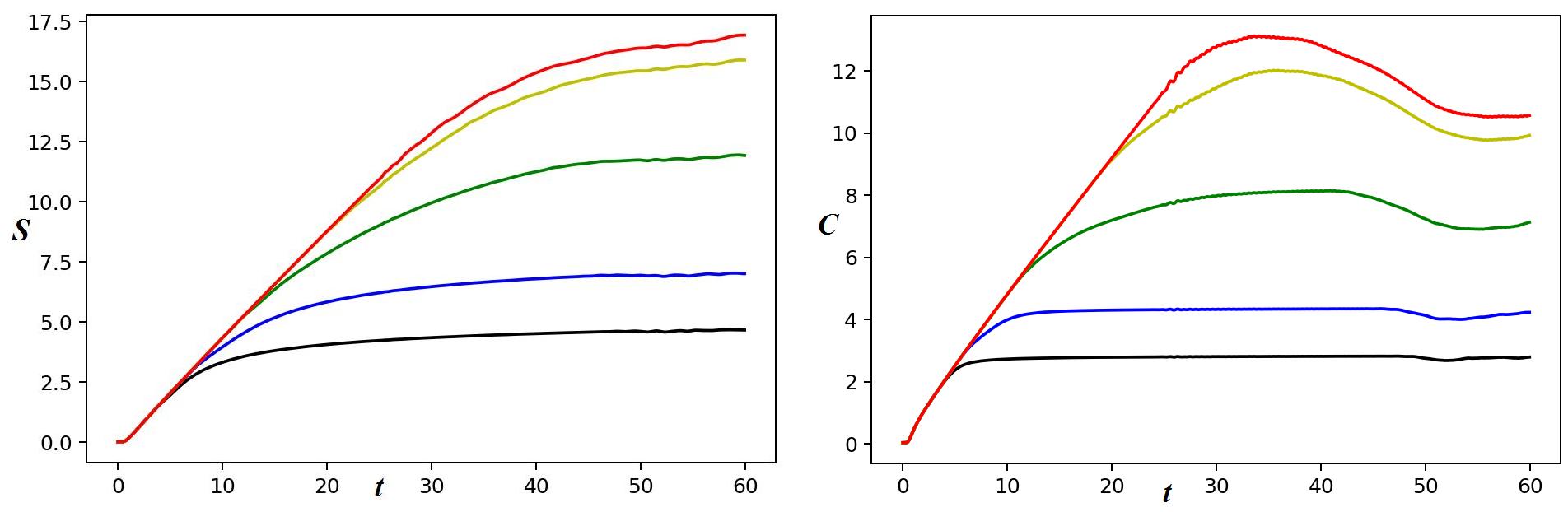}
\\
\end{center}
\caption{\small{\eqref{ee10}  quench - temporal evolution: $N=100$  and DBC ($\omega(i)=3$  and $k=1$). Lines:  black  $n=6$ ,  blue  $n=10$,  green $n=20$,  yellow  $n=36$, red    $n=50$.  The left panel  entropy. The right panel  capacity. }
\label{f7}
}
\end{figure}

\begin{figure}[!ht]
\begin{center}
\includegraphics[width=0.95\columnwidth]{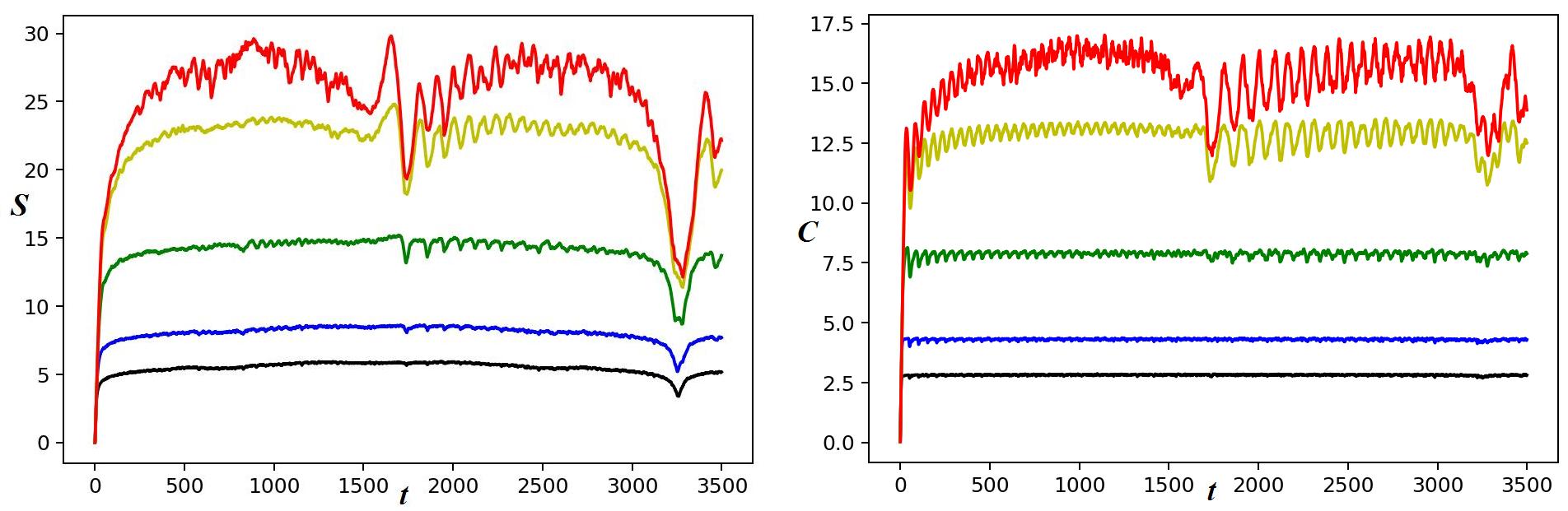}
\\
\end{center}
\caption{\small{\eqref{ee10}  quench - temporal evolution: $N=100$  and DBC ($\omega(i)=3$ and $k=1$). Lines:  black  $n=6$ ,  blue  $n=10$,  green $n=20$,  yellow  $n=36$, red    $n=50$.  The left panels  entropy. The right panels  capacity. }
\label{f8}
}
\end{figure}

 
Now,  using the results of    Sec. \ref{s3} we will  analyse the entropy and capacity of the entanglement.   To  make contact with  our previous considerations we take $\varepsilon=\sqrt{2}/3$ which corresponds to  the initial frequency $\omega(i)=3$. Let us compare it with the abrupt case.   First,  we should note that   $\omega(t)$   tends (is not equal) to zero at infinity,   so we cannot use the argument presented in Sec.  \ref{s3c}. In consequence,  it is more difficult to identify a  distinguished periodicity and the temporal evolution is more complicated  as shown in  Figs. \ref{f7} and \ref{f8}.    Despite the identical initial conditions    the  values of  the capacity are smaller than  those for the abrupt quench; moreover, the transition from the  linear  to the  oscillatory regime  is much smoother and  occurs earlier  compared to  the   abrupt quench. However, there are minima where the area law approximately  holds (especially for smaller $\omega(i)$ i.e. large $\varepsilon$). For larger times, we observe the oscillations  about  a saturation value  similarly to the abrupt case. 
\par
\begin{figure}[!ht]
\begin{center}
\includegraphics[width=0.95\columnwidth]{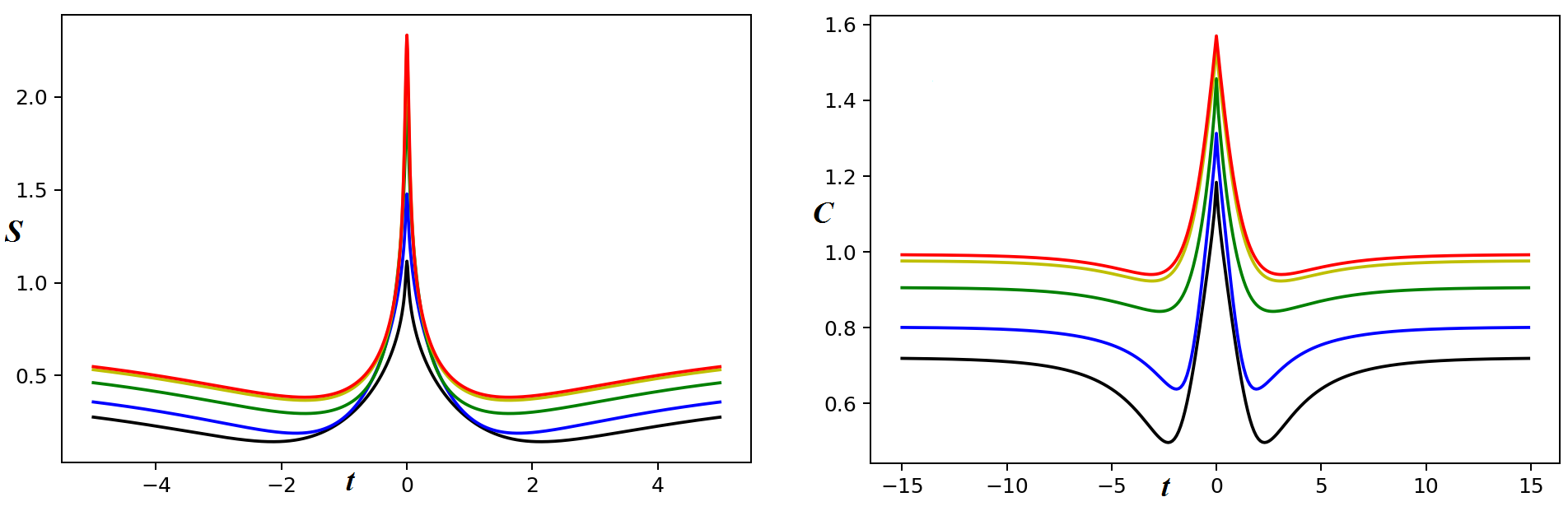} 
\\
\end{center}
\caption{\small{ \eqref{e12} quench with $\varepsilon=5$  - temporal evolution: $N=100$  and DBC ($k=1$). Lines:  black  $n=6$ ,  blue  $n=10$,  green $n=20$,  yellow  $n=36$, red    $n=50$.  The left panel  entropy. The right panel  capacity. }
\label{f9}
}
\end{figure}
\begin{figure}[!ht]
\begin{center}
\includegraphics[width=0.95\columnwidth]{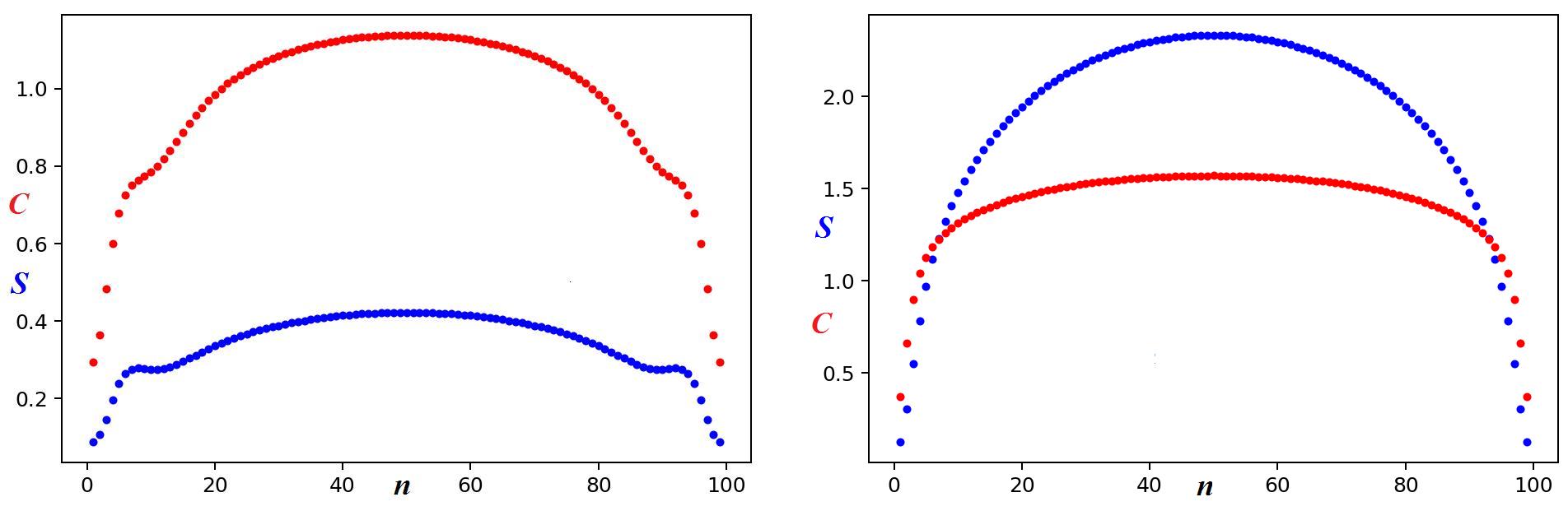}
\\
\end{center}
\caption{\small{\eqref{e12}  quench with  $\varepsilon=5$  -  time slices,  $N=100$     and DBC ($k=1$). Lines:   blue - entropy, red  - capacity. The left panel $t=1$. The right panel $t=0$.  }
}
\label{f10}
\end{figure}

 The second interesting case   is given by the initial conditions  \eqref{e15}  imposed at  $t_0=-\infty$, i.e. when the frequency is given  directly by eq. \eqref{e12}. Then, we start and end with massless (in the field  theory picture) case with the peak   $2/\varepsilon^2$ at time $t=0$. In this case the functions $b_j(t)$ are  given by
 \be
 \label{e23}
 b_j^2(t)=\frac{a_j^2\varepsilon^2+\tanh^2(t/\varepsilon)}{1+a_j^2\varepsilon^2},
 \ee
 for $j=1,\ldots, N$.
 A remarkable property of  the functions \eqref{e23}  is that they     satisfy  the condition \eqref{e15} also at $t=\infty$; there is no oscillatory behavior  for  $a_j>0$ at plus infinity (independently of the value of  parameter $\varepsilon$).  Such a property implies that we observe only  a peak  in  both the  entropy and capacity, see Fig. \ref{f9}.  More insight can be obtained when we  consider  the time-fixed  slices.  For  sufficiently large (positive and negative) time the behavior   of both the  entropy and capacity    coincide with the ones presented in Sec. \ref{s4a}. However, near $t=0$ this  behavior becomes disordered, as depicted in the left panel Fig. \ref{f10}.  At $t=0$ it  resembles the    massless case again, see  the right panel  in Fig. \ref{f10}.
\par
\begin{figure}[!ht]
\begin{center}
\includegraphics[width=0.47\columnwidth]{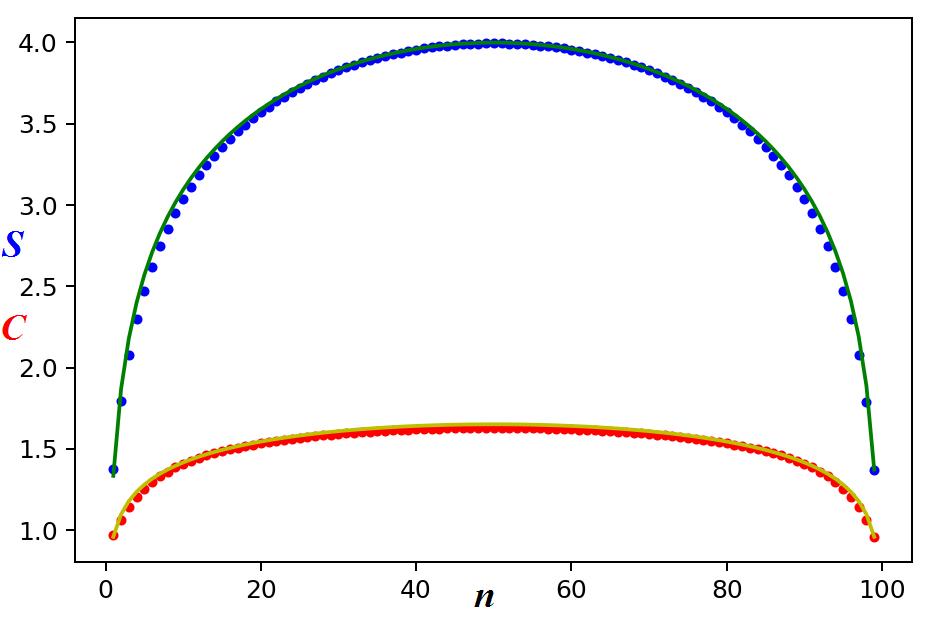}
\\
\end{center}
\caption{\small{\eqref{e12}  quench with  $\varepsilon=0.0001 $ (Dirac delta limit)  -  time slices $t=0$,  $N=100$     and DBC ($k=1$). Date points:   blue - entropy, red  - capacity.   Green line:
$ 0.77\ln((100/\pi)\sin(\pi n/100))+1.3 $, yellow line: $0.2\ln((100/\pi) \sin(\pi n/100))+0.95 $.}
\label{f11}
}
\end{figure}

 
  Finally,  as $\varepsilon\rightarrow 0$ the frequency tends to the Dirac delta. Then the  numerical results for $\varepsilon=0.0001$  and  $t\neq 0$ give the entropy and capacity   as  in Sec. \ref{s4a}, while     for  $t=0$ we  obtain     the  similar behavior; however,   with different parameters. Namely, instead  of the factor $1/6$ (see Sec. \ref{s4a})  we   have  $N$-dependent parameter;  e.g. for $N=100$ we have     $0.77$ for entropy and $0.2$  for capacity, respectively;   the suitable curves are depicted in   Fig. \ref{f11}. 

\section{Summary and final discussion} 
In this work we have   studied    the notion of the  modular entropy and capacity (in particular the  fluctuations of the entanglement entropy)   in more detail.    This is  motivated by the recent  investigations  of  both  the quantities in various physical contexts; e.g.   quantum gravitational effects. To this end  we  considered    systems  of  oscillators with  the time-dependent frequency coupled by  a  parameter.   Such models, due to the discretization procedure,  can  be used to analyse  field theory problems  resulting   from quench phenomena or  non-static gravitational metrics.   First, we showed that   the modular and   R\'enyi  entropies  have the same dynamics for the basic solutions of  the single  TDHO.  Moreover,  in  this case the capacity  is time-independent and we  found  its  form (eqs. \eqref{ee11}-\eqref{ee15}).  
\par    Next,    we studied    the aforementioned  notions for  a finite number of the TDHO   and    various   bipartite  decompositions.   As a result we obtained   analytic formulae  for them; in contrast to  the single oscillator case the dynamics of the modular entropy is different from the   R\'enyi one and the capacity becomes  time-dependent (eqs. \eqref{ee5} and \eqref{ee6}).    Next, we focused on the capacity (fluctuation)  of  the entanglement, which  has recently  gained  increasing attention.  In addition to the work  \cite{b9h} we considered  various boundary conditions  related to the  approximation  of the finite size system (in field theory picture).  Taking     $n\ll N$  and   the abrupt quenches  they coincide with the observations  made  in Ref. \cite{b9h}. Obviously, we observed  (for  $n\sim N/2$)   some differences due to  finite-size assumption   (e.g. oscillations,  in general non-linear  behavior w.r.t. the  subsystem size, and local  area laws),  see Figs. \ref{f3}-\ref{f25}.  We confirmed, in agreement with general theory,   that the capacity (as the entropy)  is  symmetric  with respect to both subsystems;  moreover,   for fixed time    they  increase with  $n$ (for   $n<N/2$).  We  analysed  also the relative fluctuation of entanglement and,  in Sec. \ref{s3c},  some conditions which give rise to   distinguished  periodicities. 
Special attention has been paid to the theoretical prediction resulting  from the quasiparticles model;  the initial   slope   coefficients, obtained in this way, are different for the  entropy and capacity and  agree  with  the ones from lattice approximation (at least  for some range of parameters).   
\par Finally, we  studied some of the above issues in the case of  continuous protocols. To this end, in Sec. \ref{s5},   we considered   frequencies which vanish  at  plus (and minus) infinity.   In particular, we  examined the model in which the frequency tends to the Dirac delta. In such a  case  the behavior of the entropy and capacity, except  one point  (time zero), coincide with the CFT results (see Sec. \ref{s4a});  fairly surprisingly,  at time zero  we obtain   similar behavior, however, with   $N$-dependent  coefficient, see Fig.  \ref{f11}. 
 \par
 All  the above  issues  have been  discussed  in the   analytical manner,  due to   the explicit forms (given by  elementary functions) of the  solutions of  the  EMP equations (numerical computations were used, for higher $N$,  to the standard matrix algebra only).  
\par
Turning to possible further developments, it would be interesting to extend the above   considerations     to  the transition from the  DBC to NBC (such a a scenario   can be used to    simulate the dynamical Casimir effect \cite{b12g}) and/or multiply quenched harmonic chains \cite{b12d}.  The other states can be also considered, e.g. the coherent or squeezed  ones, \cite{b15b,b15a}.  Additionally, some extensions   to larger dimensions are also interesting (see  \cite{b12aa,b16a} and references therein) or   higher derivative theories  \cite{b16b}.   In this context,  particularly interesting  are  applications  in field theory in expanding spacetimes \cite{b16bb,b13a}.    Finally,    basing on    the methods  presented in Refs.    \cite{b12h,b16c}, a deeper  analysis  of the capacity   (fluctuations of the  entropy)    would be worthwhile. 
\vspace{0.5cm}
\par
{\bf Acknowledgment}
\par 
The  author would like to thank    Piotr  Kosi\'nski for useful comments.

  \end{document}